\documentclass[aps,prl,preprint,tightenlines,superscriptaddress,showpacs,byrevtex]{revtex4}
\usepackage{graphicx} % Include figure files
\usepackage{epsfig}   % Include figure files ! added by T.Iijima
\usepackage{dcolumn}  % Align table columns on decimal point

\graphicspath{{ps}}

\begin{document}

%\vspace*{-3\baselineskip}
%\resizebox{!}{3cm}{\includegraphics{belle.eps}}

\preprint{\vbox{ \hbox{   }
                 \hbox{BELLE-CONF-0567}
                 \hbox{EPS05-542} 
%                 \hbox{hep-ex nnnn, if available}
}}

\title{ \quad\\[0.5cm] Measurement of Exclusive $B \to X_u \ell \nu$ Decays \\
with $D^{(*)} \ell \nu$ Decay Tagging}

%%%% >>>>> insert the authorlist here. BEFORE the abstract !!!!! <<<<<

%\collaboration{Belle Collaboration}
%\noaffiliation
%%% Paper:    
%%% Journal:  Summer 2005 conference papers
%%% Contacts: 
%%% Non-responding authors or those who said NO are commented out.
%%% ====================================================================
%%% Click the RELOAD button on your web browser to see the updated file.
%%% ====================================================================
%%% Use \input{author} to insert this material into your latex file.
%%%%% Force institutions to appear in alphabetical order when typeset.
\affiliation{Aomori University, Aomori}
\affiliation{Budker Institute of Nuclear Physics, Novosibirsk}
\affiliation{Chiba University, Chiba}
\affiliation{Chonnam National University, Kwangju}
\affiliation{University of Cincinnati, Cincinnati, Ohio 45221}
\affiliation{University of Frankfurt, Frankfurt}
\affiliation{Gyeongsang National University, Chinju}
\affiliation{University of Hawaii, Honolulu, Hawaii 96822}
\affiliation{High Energy Accelerator Research Organization (KEK), Tsukuba}
\affiliation{Hiroshima Institute of Technology, Hiroshima}
\affiliation{Institute of High Energy Physics, Chinese Academy of Sciences, Beijing}
\affiliation{Institute of High Energy Physics, Vienna}
\affiliation{Institute for Theoretical and Experimental Physics, Moscow}
\affiliation{J. Stefan Institute, Ljubljana}
\affiliation{Kanagawa University, Yokohama}
\affiliation{Korea University, Seoul}
\affiliation{Kyoto University, Kyoto}
\affiliation{Kyungpook National University, Taegu}
\affiliation{Swiss Federal Institute of Technology of Lausanne, EPFL, Lausanne}
\affiliation{University of Ljubljana, Ljubljana}
\affiliation{University of Maribor, Maribor}
\affiliation{University of Melbourne, Victoria}
\affiliation{Nagoya University, Nagoya}
\affiliation{Nara Women's University, Nara}
\affiliation{National Central University, Chung-li}
\affiliation{National Kaohsiung Normal University, Kaohsiung}
\affiliation{National United University, Miao Li}
\affiliation{Department of Physics, National Taiwan University, Taipei}
\affiliation{H. Niewodniczanski Institute of Nuclear Physics, Krakow}
\affiliation{Nippon Dental University, Niigata}
\affiliation{Niigata University, Niigata}
\affiliation{Nova Gorica Polytechnic, Nova Gorica}
\affiliation{Osaka City University, Osaka}
\affiliation{Osaka University, Osaka}
\affiliation{Panjab University, Chandigarh}
\affiliation{Peking University, Beijing}
\affiliation{Princeton University, Princeton, New Jersey 08544}
\affiliation{RIKEN BNL Research Center, Upton, New York 11973}
\affiliation{Saga University, Saga}
\affiliation{University of Science and Technology of China, Hefei}
\affiliation{Seoul National University, Seoul}
\affiliation{Shinshu University, Nagano}
\affiliation{Sungkyunkwan University, Suwon}
\affiliation{University of Sydney, Sydney NSW}
\affiliation{Tata Institute of Fundamental Research, Bombay}
\affiliation{Toho University, Funabashi}
\affiliation{Tohoku Gakuin University, Tagajo}
\affiliation{Tohoku University, Sendai}
\affiliation{Department of Physics, University of Tokyo, Tokyo}
\affiliation{Tokyo Institute of Technology, Tokyo}
\affiliation{Tokyo Metropolitan University, Tokyo}
\affiliation{Tokyo University of Agriculture and Technology, Tokyo}
\affiliation{Toyama National College of Maritime Technology, Toyama}
\affiliation{University of Tsukuba, Tsukuba}
\affiliation{Utkal University, Bhubaneswer}
\affiliation{Virginia Polytechnic Institute and State University, Blacksburg, Virginia 24061}
\affiliation{Yonsei University, Seoul}
  \author{K.~Abe}\affiliation{High Energy Accelerator Research Organization (KEK), Tsukuba} % KEK
  \author{K.~Abe}\affiliation{Tohoku Gakuin University, Tagajo} % TohokuGakuin
  \author{I.~Adachi}\affiliation{High Energy Accelerator Research Organization (KEK), Tsukuba} % KEK
  \author{H.~Aihara}\affiliation{Department of Physics, University of Tokyo, Tokyo} % Tokyo
  \author{K.~Aoki}\affiliation{Nagoya University, Nagoya} % Nagoya
  \author{K.~Arinstein}\affiliation{Budker Institute of Nuclear Physics, Novosibirsk} % BINP
  \author{Y.~Asano}\affiliation{University of Tsukuba, Tsukuba} % Tsukuba
  \author{T.~Aso}\affiliation{Toyama National College of Maritime Technology, Toyama} % Toyama
  \author{V.~Aulchenko}\affiliation{Budker Institute of Nuclear Physics, Novosibirsk} % BINP
  \author{T.~Aushev}\affiliation{Institute for Theoretical and Experimental Physics, Moscow} % ITEP
% ...
% Lausanne
\collaboration{The Belle Collaboration}

\vspace{1.0cm}

\begin{abstract}
We report a measurement of the charmless semileptonic decays 
$B^0 \to \pi^- / \rho^- \ell^{+} \nu$ and $B^+ \to \pi^0 / \rho^0 \ell^{+} \nu$,  
based on 253 fb$^{-1}$ of data collected with the Belle detector at the KEKB $e^+e^-$ asymmetric collider.
In this analysis, the accompanying $B$ meson is reconstructed in the
semileptonic mode $B \rightarrow D^{(*)} \ell \nu$, enabling detection of the signal modes with high 
purity.
%and with marginal statistics.
We measure the branching fractions
${\cal B}(B^0 \to \pi^- \ell^+ \nu)  = 
(1.48\pm 0.20\pm 0.16\pm 0.04) \times 10^{-4}$,
${\cal B}(B^0 \to \rho^- \ell^+ \nu)  = 
(2.07\pm 0.47\pm 0.25\pm 0.14) \times 10^{-4}$,
${\cal B}(B^+ \to \pi^0 \ell^+ \nu)  = 
(0.76\pm 0.13\pm 0.08\pm 0.04) \times 10^{-4}$ and
${\cal B}(B^+ \to \rho^0 \ell^+ \nu)  = 
(1.39\pm 0.23\pm 0.17\pm 0.03) \times 10^{-4}$,
where the errors are statistical, experimental systematic, and systematic due
to form-factor uncertainties, respectively.
For each mode we also present the partial branching fractions in three $q^2$ intervals; 
$q^2 < 8$, $8 \leq q^2 < 16$, and $q^2 \geq 16$\,GeV$^2/c^2$.
Based on these results, the magnitude of the Cabibbo-Kobayashi-Maskawa matrix 
element $V_{ub}$ is extracted. 
All of the presented results are preliminary.
\end{abstract}


\maketitle

%%%% >>>> keep the final version single-spaced
\tighten

{\renewcommand{\thefootnote}{\fnsymbol{footnote}}}
\setcounter{footnote}{0}

%%%--------------------------------------------------------------------------%%%
\section{Introduction}
\label{sec:Introduction}
Exclusive $B \to X_u \ell \nu$ decays proceed dominantly via a $b \to u W^-$ tree 
process 
%with a spectator quark 
and can be used to determine $|V_{ub}|$, 
one of the smallest and least known elements of the 
Cabibbo-Kobayashi-Maskawa matrix~\cite{KM}.
A major theoretical uncertainty stems from the specification of form-factors (FF)  involved in the decays. 
The recent release of preliminary results from unquenched Lattice QCD(LQCD) calculations of FFs~\cite{Shigemitsu,Okamoto} has made possible the 
model-independent determination of $|V_{ub}|$. 
Since LQCD results are available only in the high $q^2$ region ($\geq 16$\,GeV$^2/c^2$), a clean measurement of the partial $B \to \pi \ell \nu$ branching fraction in the same high $q^2$ region is essential.

There have been several measurements in the past by CLEO, BaBar and
Belle for the $B \to \pi \ell \nu$, $\rho \ell \nu$, $\eta \ell \nu$
and $\omega \ell \nu$ modes.
The analyses in ~\cite{CLEO1996,CLEO2000,CLEO2003,BABAR2003,BELLE2004,
BABAR2005} utilize the method, originally developed by CLEO, where the 
$B$ decays are reconstructed by inferring the undetected neutrino 
momentum from missing energy and momentum (``$\nu$-reconstruction 
method'')~\cite{CLEO1996}.
In the $B$-factory era, we may straightforwardly improve the statistical 
precision by simply applying the $\nu$-reconstruction method.
However, the systematic uncertainty limits the measurement because of a poor signal-to-noise ratio.

The Belle collaboration has previously presented measurements of $B^0 \to \pi^- / 
\rho^- \ell^+ \nu$ via $D^{(*)} \ell \nu$ decay tagging based on
140 fb$^{-1}$ data~\cite{BELLE_SLTAG}.
The work demonstrated that the method allows one to detect the 
signal with high purity and is a promising approach to yield the
best overall precision from the large data sample being accumulated 
by $B$-factories over the next few years.
The BaBar collaboration also presented measurements of exclusive
$B \to X_u \ell \nu$ decays, both with $D^{(*)} \ell \nu$ decay tagging
~\cite{BABAR_SLTAG1, BABAR_SLTAG2} and with hadronic decay tagging
~\cite{BABAR_FLTAG}.

In this paper we present measurements of $B^0 \to \pi^- / \rho^- \ell^+ \nu$ 
and $B^+ \to \pi^0 / \rho^0 \ell^+ \nu$ decays using $D^{(*)} \ell \nu$ decay 
tagging and based on 253 fb$^{-1}$ data.
In comparison with the previous measurement~\cite{BELLE_SLTAG}, the analysis has been 
extended to higher integrated luminosity and to the additional modes
$B^+ \to \pi^0 / \rho^0 \ell^+ \nu$.
The analysis technique is nearly identical to the previous one;
we reconstruct the entire decay chain from the $\Upsilon(4S)$,
$\Upsilon(4S) \to B_{sig}B_{tag}$, $B_{sig} \to \pi / \rho \ell \nu$ and
$B_{tag} \to D^{(*)} \ell \bar{\nu}$ with several $D^{(*)}$ sub-modes.
The back-to-back correlation of the two $B$ mesons in the $\Upsilon(4S)$
rest frame allows us to constrain the kinematics of the double semileptonic 
decay.
%We present the simultaneous extraction of $B^0 \to \pi^- / \rho^- \ell^+ \nu$ 
%and $B^+ \to \pi^0 / \rho^0 \ell^+ \nu$.
Inclusion of charge conjugate decays is implied throughout this paper.

%%%--------------------------------------------------------------------------%%%
\section{Data Set and Experiment}
\label{sec:data_exp}

The analysis is based on data recorded with the Belle detector at the 
asymmetric $e^+e^-$ collider KEKB operating at the center-of-mass (c.m.) 
energy of the $\Upsilon(4S)$ resonance~\cite{KEKB}.
The $\Upsilon(4S)$ data set used for this study corresponds to an
integrated luminosity of 253 fb$^{-1}$ and contains $275.2 \times 10^6$ 
$B \bar{B}$ events.

The Belle detector is a large-solid-angle magnetic spectrometer
that consists of a silicon vertex detector (SVD),
a 50-layer central drift chamber (CDC), 
an array of aerogel threshold \v{C}erenkov counters (ACC),
a barrel-like arrangement of time-of-flight scintillation counters (TOF),
and an electromagnetic calorimeter comprised of CsI(Tl) crystals (ECL)
located inside a super-conducting solenoid coil 
that provides a 1.5~T magnetic field.  
An iron flux-return located outside of the coil is instrumented
to detect $K_L^0$ mesons and to identify muons (KLM).  
The detector is described in detail elsewhere~\cite{BELLE}. 
Two inner detector configurations were used. A 2.0 cm beampipe
and a 3-layer silicon vertex detector was used for the first sample
of 152 million $B\bar{B}$ pairs, while a 1.5 cm beampipe, a 4-layer
silicon detector, and a small-cell inner drift chamber were used to record  
the remaining 123 million $B\bar{B}$ pairs\cite{Ushiroda}.  

A detailed Monte Carlo (MC) simulation, which fully describes the detector
geometry and response and is based on GEANT~\cite{GEANT}, is
applied to estimate the signal detection efficiency and to study the 
background.
To examine the FF dependence, MC samples for the 
$B \to \pi \ell \nu$ signal decays are generated 
with different form-factor models;
a quark model (ISGW~II ~\cite{ISGW2}), 
light cone sum rules (LCSR; Ball'01 for $\pi \ell \nu$ ~\cite{Ball01} 
and Ball'98 for $\rho \ell \nu$ ~\cite{Ball98}) and 
quenched lattice QCD (UKQCD ~\cite{UKQCD}).
A relativistic quark model (Melikhov ~\cite{Melikhov}) is also 
used for $\rho \ell \nu$.
To model the cross-feed from other $B \to X_u \ell \nu$ decays,
MC samples are generated with the ISGW~II model for the resonant 
components and the DeFazio-Neubert model ~\cite{Fazio-Neubert} for 
non-resonant component ($\pi \ell \nu$ and $\rho \ell \nu$ 
components are excluded in this sample).
To model the $B\bar{B}$ and continuum backgrounds, large generic 
$B\bar{B}$ (based on Evtgen~\cite{Evtgen}) and $q\bar{q}$ Monte Carlo 
samples are used.

%%%--------------------------------------------------------------------------%%%
\section{Event Reconstruction and Selection}

%\subsection{Hadron Tracks}
Charged particle tracks are reconstructed from hits in the SVD and CDC. 
They are required to satisfy track quality cuts based on their impact 
parameters relative to the measured profile of the interaction point 
(IP profile) of the two beams. 
Charged kaons are identified by combining information on ionization loss 
($dE/dx$) in the CDC, \v{C}herenkov light yields in the ACC and time-of-flight 
measured by the TOF system.
For the nominal requirement, the kaon identification efficiency is 
approximately $88\%$ and the rate for misidentification of pions as 
kaons is about $8\%$.
Hadron tracks that are not identified as kaons are treated as pions.
Tracks satisfying the lepton identification criteria, as described later, 
are removed from consideration.

Neutral pions are reconstructed using $\gamma$ pairs with an invariant mass 
between 117 and 150\,MeV/$c^2$.
Each $\gamma$ is required to have a minimum energy deposit of 
$50$~MeV.
$K_S^0$ mesons are reconstructed using pairs of charged tracks that
have an invariant mass within $\pm 12$\,MeV/$c^2$ of the known $K_S^0$ mass.

%\subsection{Leptons}
Electron identification is based on a combination of $dE/dx$ in CDC,
the response of ACC, shower shape in ECL and the ratio of energy 
deposit in ECL to the momentum measured by the tracking system.
Muon identification by KLM is performed by resistive plate counters 
interleaved in the iron yoke.
The lepton identification efficiencies are estimated to be about 90\% 
for both electrons and muons in the momentum region above 1.2\,GeV/$c$. 
where leptons from prompt $B$ decays dominate.
The hadron misidentification rate is measured 
using reconstructed $K_S^0 \to \pi^+ \pi^-$
and found to be less than 0.2\% for electrons and 1.5\% for muons
in the same momentum region.

For the reconstruction of $B_{tag} \to D^{(*)} \ell \bar{\nu}$, 
the lepton candidate is required to have the correct sign charge with
respect to the $D$ meson flavor and a laboratory momentum greater
than 1.0\,GeV/$c$ ($p_{\ell}^{lab} > 1.0$ GeV/$c$). 
The $D$ meson candidates are reconstructed by using 7 decay modes of $D^+$ --
$D^+ \to K^- \pi^+ \pi^+$, $K^- \pi^+ \pi^+ \pi^0$, $K_S^0 \pi^+$, 
$K_S^0 \pi^+ \pi^0$, $K_S^0 \pi^+ \pi^+ \pi^-$,
$K^+ K_S^0$, $K^+ K^- \pi^+$-- and  10 decay modes of $D^0$ -- 
$D^0 \to K^- \pi^+$, $K^- \pi^+ \pi^0$, $K^- \pi^+ \pi^+ \pi^-$, $K_S^0 \pi^0$, 
$K_S^0 \pi^+ \pi^-$, $K_S^0 \pi^+ \pi^- \pi^0$, $K^- \pi^+ \pi^+ \pi^- \pi^0$,
$K^+ K^-$, $K_S^0 K^+ K^- $, $K_S^0 K^- \pi^+$.
The candidates are required to have an invariant mass $m_D$ within
$\pm 2.5\sigma$ ($\sigma$ is a standard deviation) of the nominal $D$ mass, 
where the mass resolution $\sigma$ is dependent on the decay mode. 
$D^{*}$ mesons are reconstructed by combining the $D$ meson candidate and 
a charged or neutral pion, 
$D^{*+} \to D^0 \pi^+$, $D^+ \pi^0$ and $D^{*0} \to D^0 \pi^0$.
Each $D^*$ candidate is required to have a mass difference 
$\Delta m = m_{\bar{D}\pi} - m_{\bar{D}}$ 
within $\pm 2.5 \sigma$ of the nominal values.    

%\subsection{$D^{(*)}$ Mesons}

%\subsection{$X_u$ System}
For the reconstruction of $B_{sig} \to X_u \ell \nu$, 
the lepton candidate is required to have the right sign charge with
respect to the $X_u$ system and $p_{\ell}^{lab} > 0.8$ GeV/$c$.
The $X_u$ system may consist of one pion or two pions (
$N_{\pi^+}=1$ or $N_{\pi^+}=N_{\pi^0}=1$ for a $\bar{B^0}$ tag and 
$N_{\pi^0}=1$ or $N_{\pi^+}=N_{\pi^-}=1$ for a $B^-$ tag).
The event is required to have no additional charged tracks or $\pi^0$ 
candidates.
We also require that the residual energy from neutral clusters be
less than 0.15 GeV ($E_{neut} < 0.15$\,GeV).   
The two leptons on the tag and the signal sides are required to have 
opposite charge.
The loss of signal due to $B^0 - \bar{B^0}$ mixing is estimated by 
MC simulation.

\begin{figure}[htbp]
 \begin{center}
  \mbox{\psfig{figure=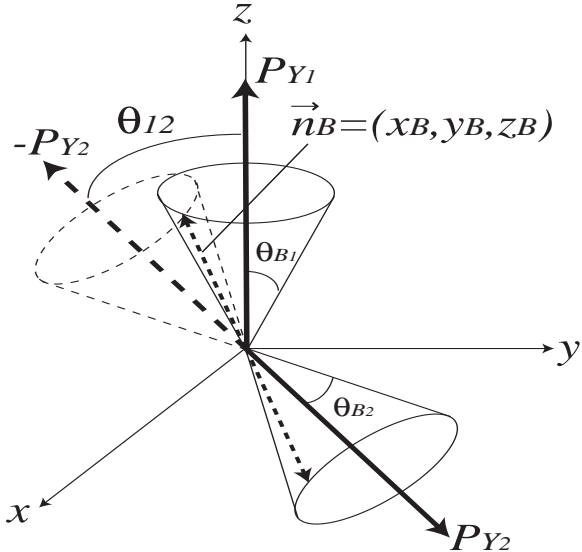,width=2.5in, height=2.5in, angle=0, scale=1.2 } } 
  \caption{Kinematics of the double semileptonic decay.}
  \label{fig:double_cone}
  \end{center}
\end{figure}

%\subsection{Event Kinematics}
We then impose a constraint based on the kinematics of the double semileptonic decay
in the $\Upsilon(4S)$ rest frame.
In the semileptonic decay on each side, $B_{1(2)} \to Y_{1(2)} \nu$ 
($Y_1 = D^{(*)} \ell$ and $Y_2 = X_u \ell$), the angle between the 
$B_{1(2)}$ meson and the detected $Y_{1(2)}$ system 
$\theta_{B_{1(2)}}$ is calculated from the relation,
$p_{\nu}^2 = (p_{B} - p_{Y})^2 = 0$ and the known $P_B$ (the absolute momentum 
of the mother $B$ meson).
This means that the $B_{1(2)}$ direction is constrained on the surface of a 
cone defined with the angle $\theta_{B_{1(2)}}$ around the direction 
of the $Y_{1(2)}$ system, as shown graphically in 
Figure~\ref{fig:double_cone}.
Then the back-to-back relation of the two $B$ meson directions implies that
the real $B$ direction is on the cross lines of the two cones when one of
the $B$ system is spatially inverted.
Denoting $\theta_{12}$ the angle between the $D^{*} \ell$ and the 
$X_u \ell$ systems, the $B$ directional vector $\vec{n}_B = (x_B, y_B, z_B)$ 
is given by, 
$z_B = \mbox{cos}\theta_{B_1}$,  
$y_B = (\mbox{cos}\theta_{B_2} - \mbox{cos}\theta_{B_2}\mbox{cos}\theta_{12})
     / \mbox{sin}\theta_{12}$,
and
\begin{eqnarray}
 x_B = \pm \sqrt{1-\frac{1}{\mbox{sin}\theta_{12}}
(\mbox{cos}^2\theta_{B_1} + \mbox{cos}^2\theta_{B_2} 
- 2 \mbox{cos}\theta_{B_1}\mbox{cos}\theta_{B_2}\mbox{cos}\theta_{12})}
 \label{eq:x_B}
\end{eqnarray}
with the coordinate definition in Figure~\ref{fig:double_cone}.
If the hypothesis of the double semileptonic decay is correct and all
the decay products are detected except for the two neutrinos, $x_B^2$
must range from 0 to 1.
Events passing a rather loose cut $x_B^2 > -2.0$ are used for signal 
extraction at a later stage of the analysis.

%%%%%%%%%%%%%%%%%%%%%%%%%%%%%%
%%%%%%%%%%%%%%%%%%%%%%%%%%%%%%

Since Eq.~\ref{eq:x_B} has two solutions and the direction of the $B$ meson is 
not uniquely determined,
we calculate, $q^2$ as $q^{2} = (E^{*}_{beam} - E^{*}_{X_u})^2 - p^{*}_{X_u}{^2}$,
using the beam energy ($E^*_{beam}$), energy ($E^*_{X_u}$) and 
momentum ($p^*_{X_u}$) of the $X_u$ system and neglecting the momentum of the 
$B$ meson in the c.m. system.
The signal Monte Carlo simulation predicts the average $q^2$ resolution
to be approximately 0.75\,GeV$^2/c^2$.

%\subsection{Background Study}
According to Monte Carlo simulations, the major backgrounds originate with 
$B \to X_c \ell \nu$ and non-signal $B \to X_u \ell \nu$ decays, 
where some particles escape detection.
There are sizable contributions from cross talk between the $\bar{B^0}$ and
$B^+$ tags.
The contribution from $q\bar{q}$ processes is found to be negligible.

%\subsection{Signal Efficiency}
For events selected as described above, the signal MC simulation 
indicates that the total detection efficiency ($\epsilon_{total}$), on the 
average of the electron and muon channels, is
$2.00\times 10^{-3}$ for $\pi^- \ell^+ \nu$ and
$7.75\times 10^{-4}$ for $\rho^- \ell^+ \nu$,
$1.49\times 10^{-3}$ for $\pi^0 \ell^+ \nu$ and
$1.76\times 10^{-3}$ for $\rho^0 \ell^+ \nu$
assuming the LCSR model.
Here, $\epsilon_{total}$ is defined with respect to the number of 
produced $B \bar{B}$ pairs where one $B$ decays into the signal mode, and 
includes the loss of signal due to $B^0 - \bar{B^0}$ mixing.
Because of the relaxed lepton momentum cut ($>0.8$\,GeV/$c$), the variation 
of efficiency with different FF models is relatively small.

\begin{figure}[htbp]
\vspace{1.0cm}
  \begin{center}
    \mbox{\psfig{figure=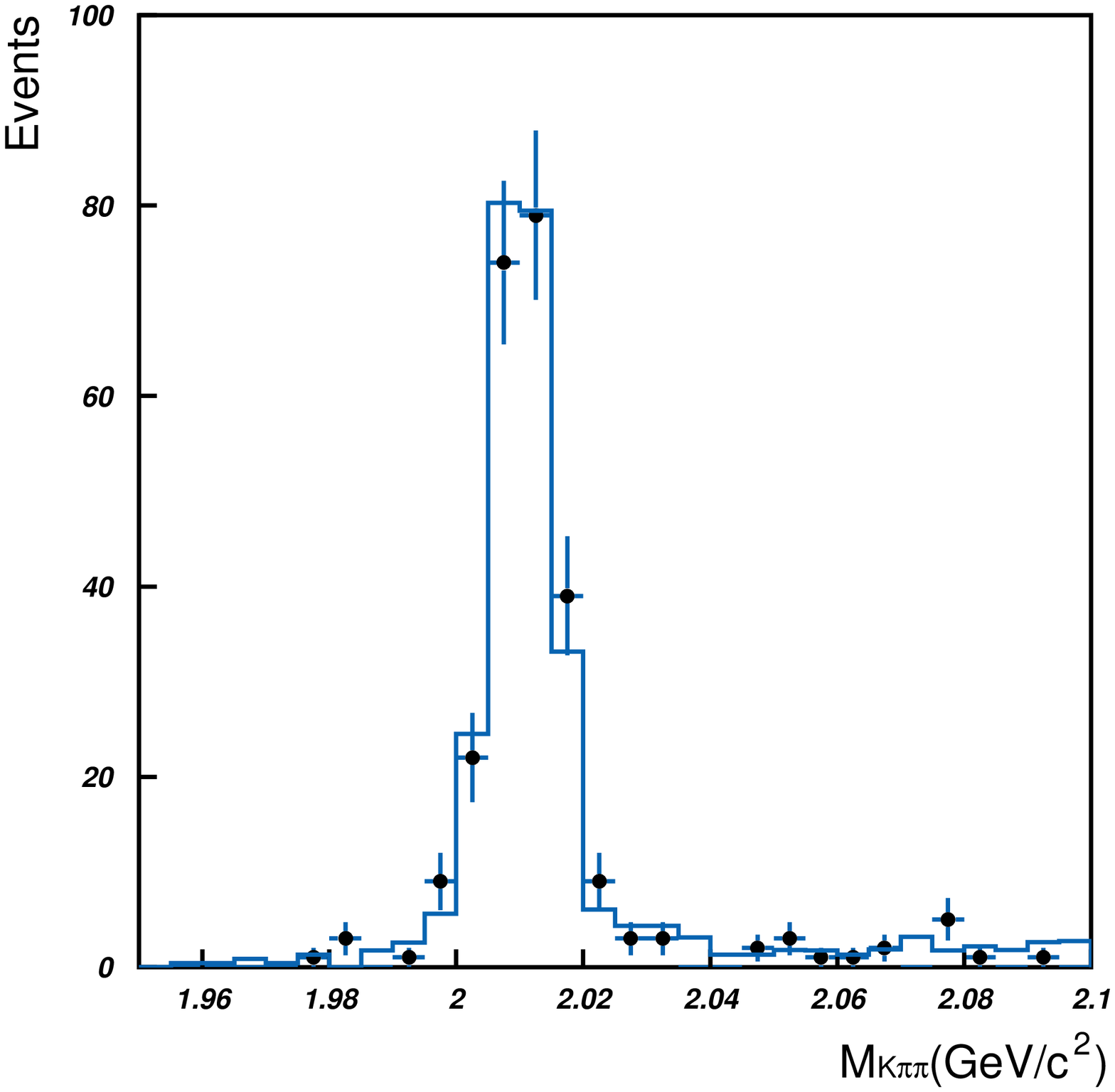,width=2.5in, height=2.5in, angle=0, scale=1. } }
    \mbox{\psfig{figure=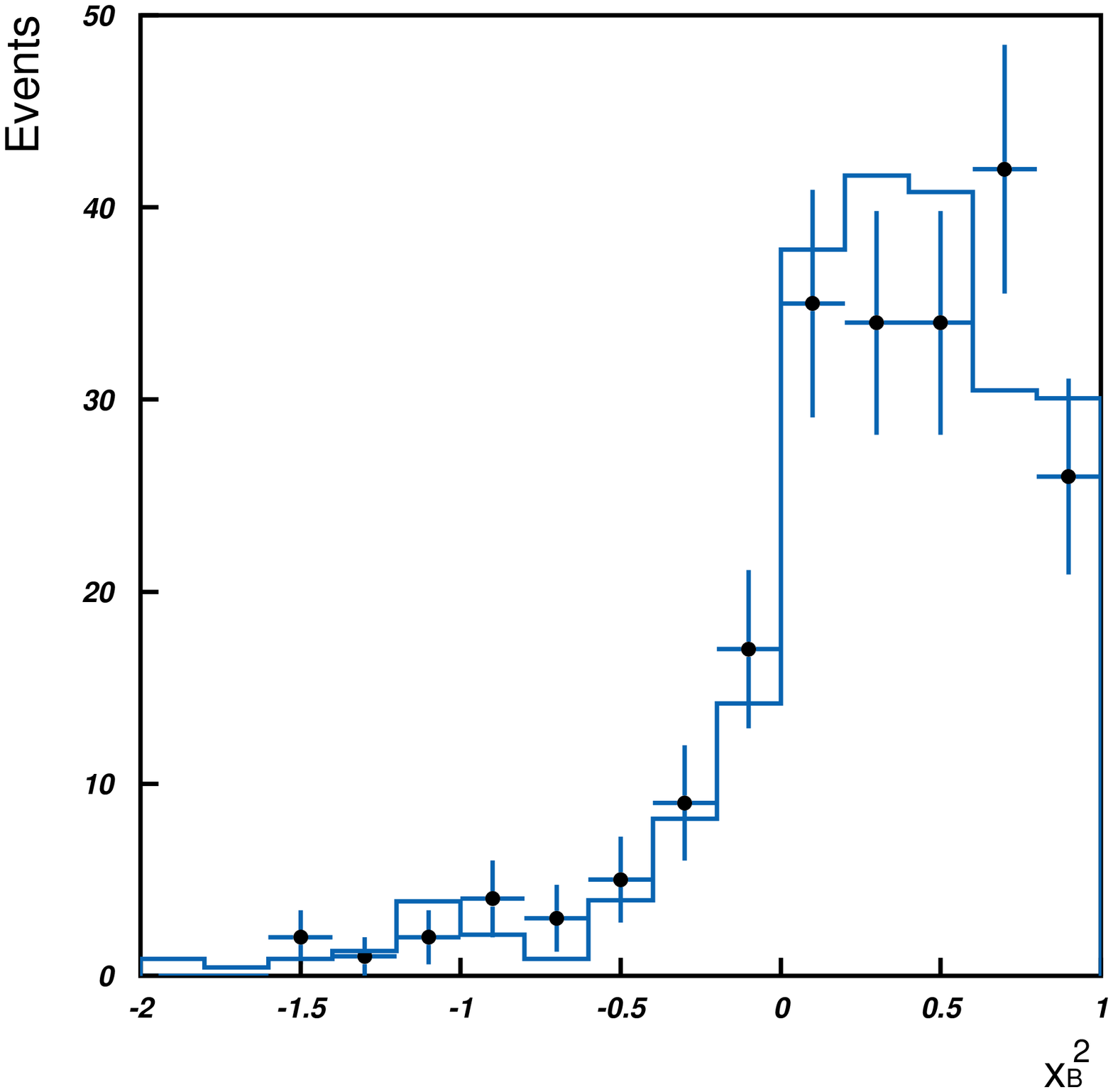,width=2.5in, height=2.5in, angle=0, scale=1. } }
    \caption{ Reconstructed $M(K\pi\pi)$ distribution(left) and $x^2_B$ distribution(right) 
   for the $D^{*-} \ell^{+} \nu$ calibration decay. Points with error bars are data and the
   histogram is the signal MC.}
    \label{fig:dstlnu}
  \end{center}
\end{figure}

%\section{Calibration using $B^0 \to D^{*-} \ell^{+} \nu$ Sample}
The validity of the method to reconstruct the double semileptonic decay 
is checked by reconstructing the signal as $B_{sig}^0 \to D^{*-} \ell^{+} \nu$ followed 
by $D^{*-} \to \bar{D^0} \pi^{-}, \bar{D^0} \to K^{+} \pi^{-}$, with 
the same requirement on the tagging side.
Figure~\ref{fig:dstlnu}-a) shows the obtained $M_{K\pi\pi}$ distribution
and its MC expectation.
With the 253 fb$^{-1}$ data sample, $224.7 \pm 15.4$ decays are clearly 
identified, while the expectation  based on the 
MC efficiency and the product branching fraction 
${\cal B}(B^0 \to D^{*-} \ell^{+} \nu, D^{*-} \to \bar{D^0} \pi^{-},
\bar{D^0} \to K^{+} \pi^{-}) = (1.38 \pm 0.06) \times 10^{-3}$~\cite{PDG2005} is $224.5 \pm 9.5$ events.
Their ratio $R=1.001 \pm 0.093$ is consistent with unity within the 
error, where 8.3\% originates from the statistics of the detected 
$D^{*-} \ell^+ \nu$ decays and 4.9\% from the error on 
${\cal B}(B^0 \to D^{*-} \ell^+ \nu)$.
Figure~\ref{fig:dstlnu}-b) shows the comparison of the reconstructed 
$x_B^2$ distribution using the same sample to the MC simulation.
The agreement between the data and MC demonstrates the validity of the
present measurement.

%%%%%%%%%%%%%%%%%%%%
%%%%%%%%%%%%%%%%%%%%

\section{Extraction of Branching Fractions}
\label{sec:SignalExtraction}
%\subsection{Fitting Method}
The $B^0 \to \pi^- / \rho^- \ell^+ \nu$ and 
$B^+ \to \pi^0 / \rho^0 \ell^+ \nu$ signals 
are extracted by fitting the obtained two-dimensional distribution in 
$(x_B^2, M_X)$, where $M_X$ is the invariant mass of the $X_u$ system.
Here, the fit components are the four signal modes, 
the other $B^0 \to X_u^- \ell^+ \nu$ and $B^+ \to X_u^0 \ell^+ \nu$
backgrounds and the $B \bar{B}$ background (seven components in total).
The PDF (probability distribution function) for each fit component 
is determined from MC simulation. 
The $\pi / \rho\, \ell^+ \nu$ signal events exhibit characteristic behavior in 
both of their $x_B^2$ and $M_X$ distributions; other $B \to X_u \ell^+ \nu$ 
events exhibit a weak peaking structure in $x_B^2$ but a broad distribution in 
$M_X$; the $B \bar{B}$ background has a relatively flat distribution 
in $x_B^2$ and a broad structure in $M_X$. 
The PDFs in $(x_B^2, M_X)$ for each of the seven fit components are prepared for both  $\bar{B^0}$ and $B^-$ tag candidates.
We then fit the two $(x_B^2, M_X)$ distributions simultaneously, constraining contributions from the cross talk between 
$\bar{B^0}$ and $B^-$ tags as well as the cross-feed between $\pi \ell^+ \nu$
and $\rho \ell^+ \nu$. 
The fitting is constrained so that the sum of the deduced branching fractions 
for $B \to \pi \ell^+ \nu$, $B \to \rho \ell^+ \nu$ and $B \to$ other 
$X_u \ell^+ \nu$ is equal to the total inclusive branching fraction 
${\cal B}(B \to X_u \ell \nu) = (0.25 \pm 0.06)$\%~\cite{BELLE_Kakuno}.

\begin{figure}[htbp]
 \begin{center}
  \begin{tabular}{ccc}
   \hspace{-0.0cm}{\mbox{\psfig{figure=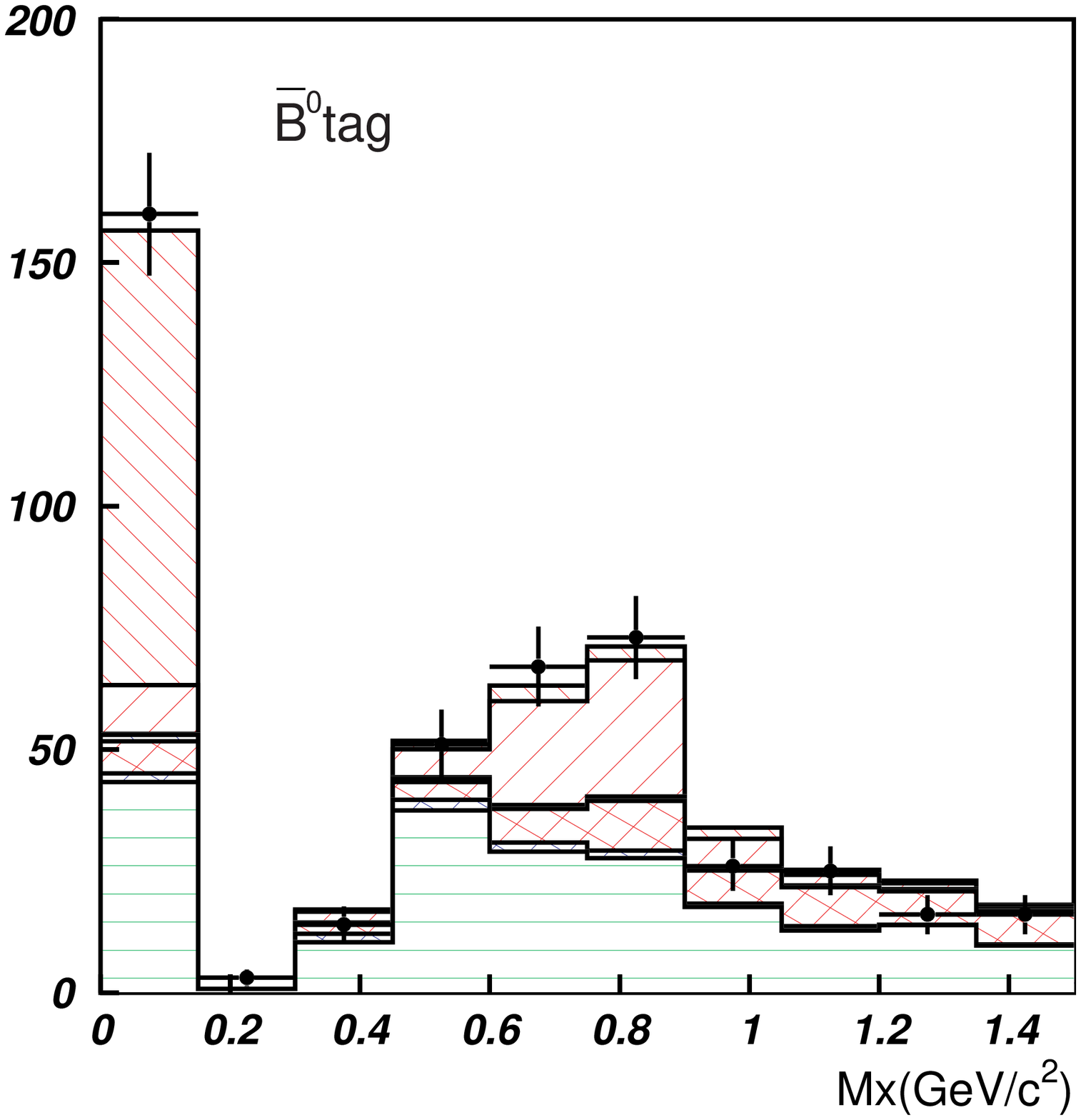,width=2.0in, height=2.0in, angle=0, scale=1.0 }} } &
   \hspace{-0.0cm}{\mbox{\psfig{figure=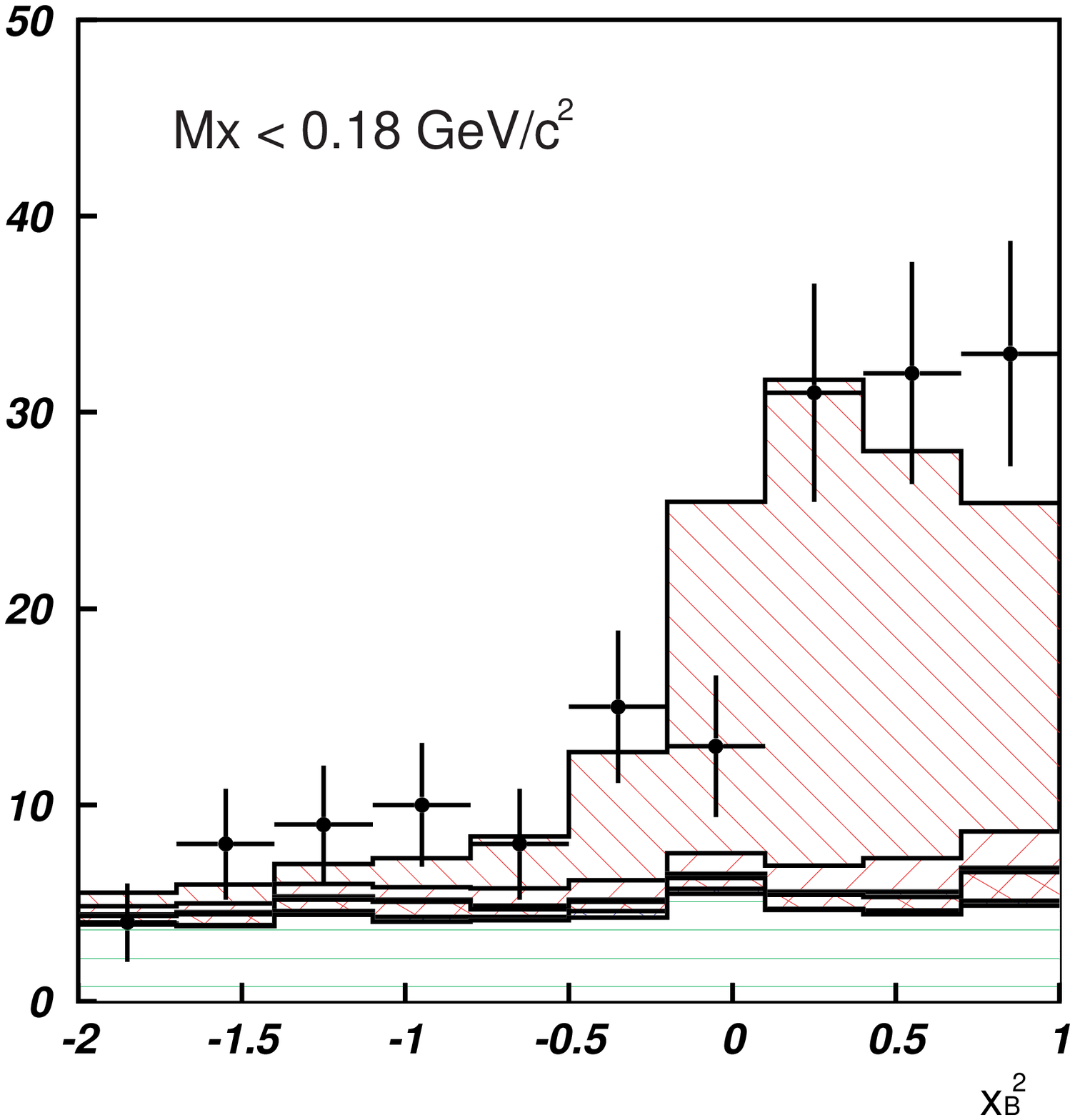,width=2.0in, height=2.0in, angle=0, scale=1.0 } }} &
   \hspace{-0.0cm}{\mbox{\psfig{figure=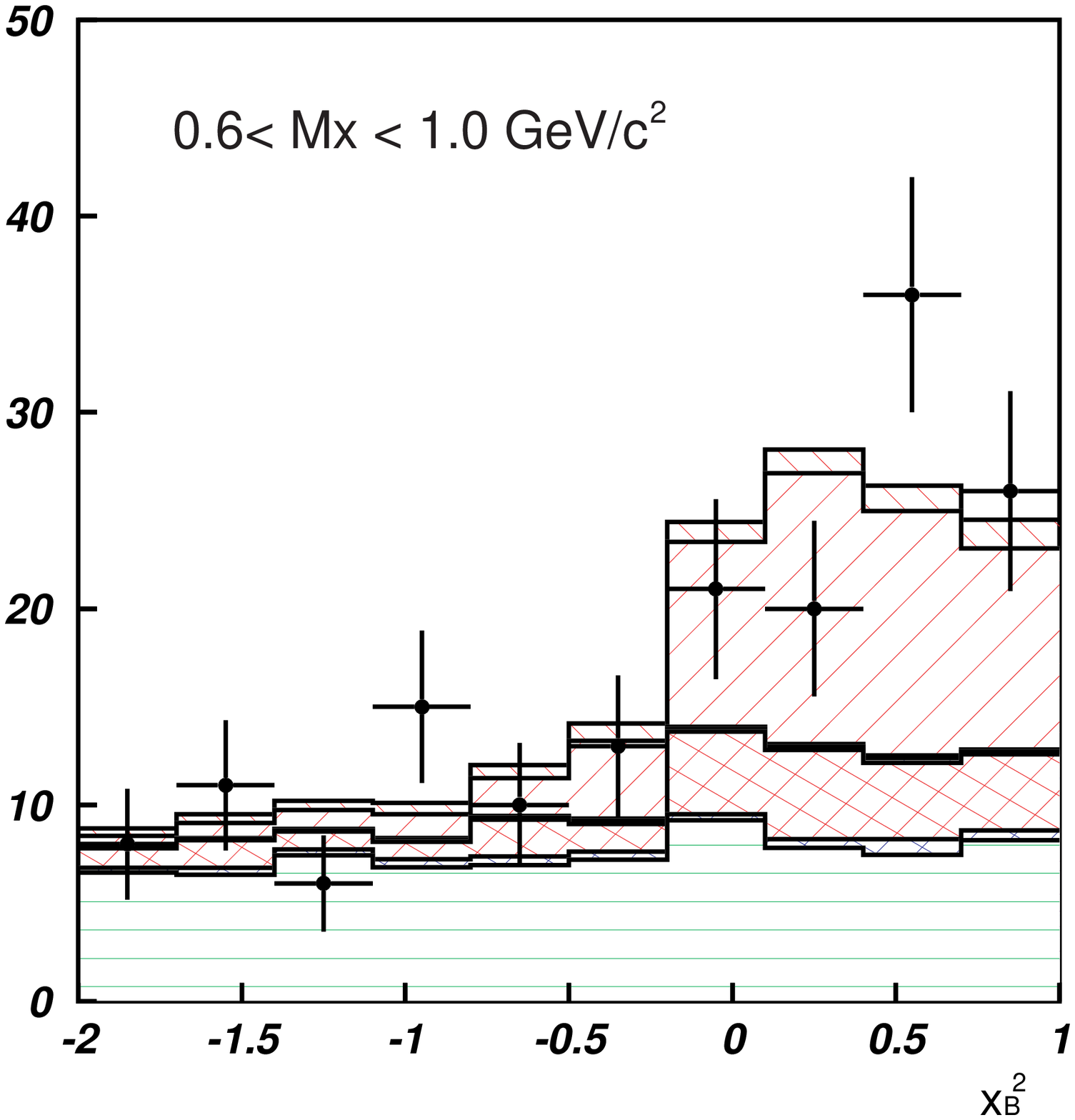,width=2.0in, height=2.0in, angle=0, scale=1.0 }} } \\

   \hspace{-0.0cm}{\mbox{\psfig{figure=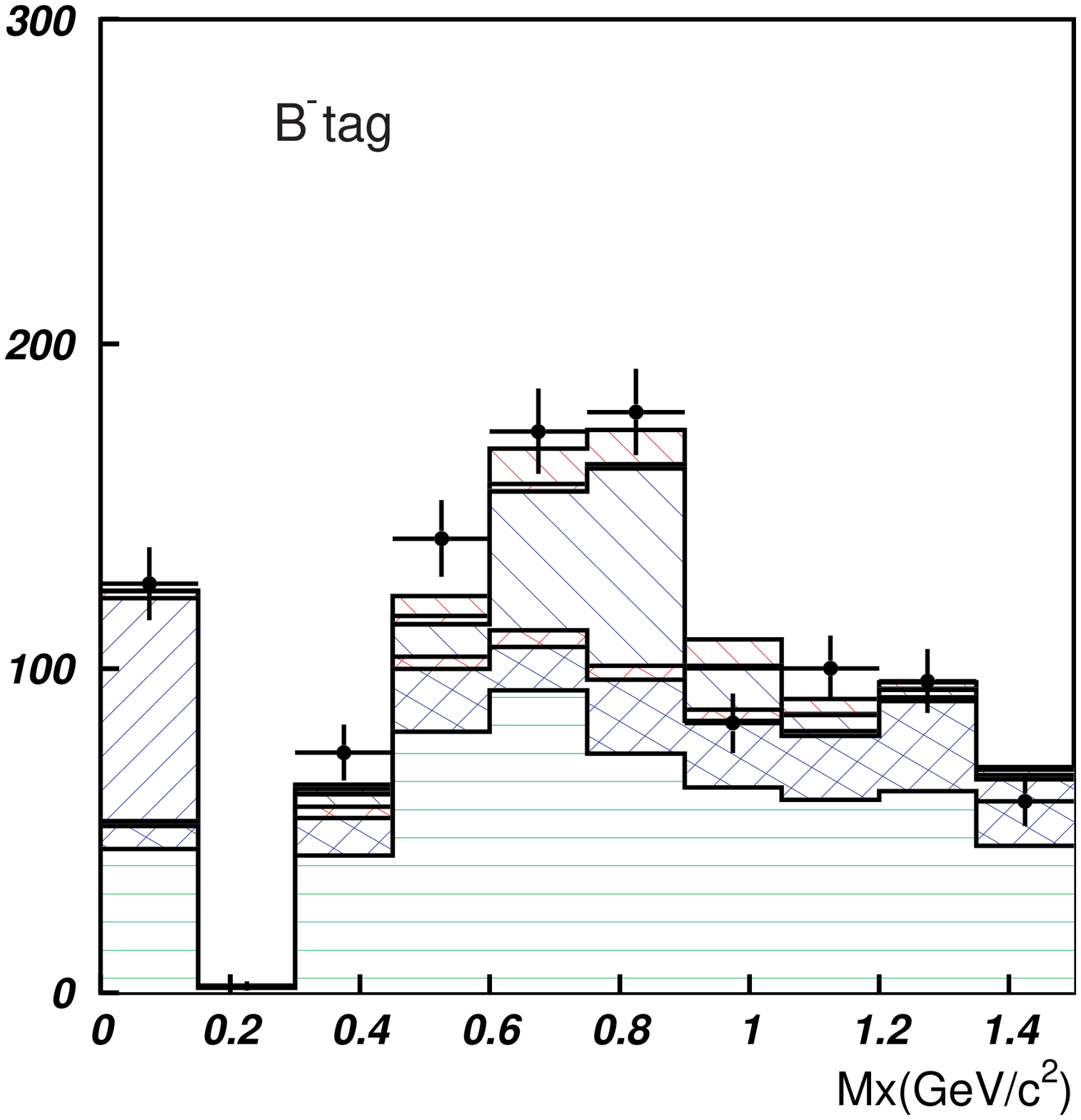,width=2.0in, height=2.0in, angle=0, scale=1.0 }} } &
   \hspace{-0.0cm}{\mbox{\psfig{figure=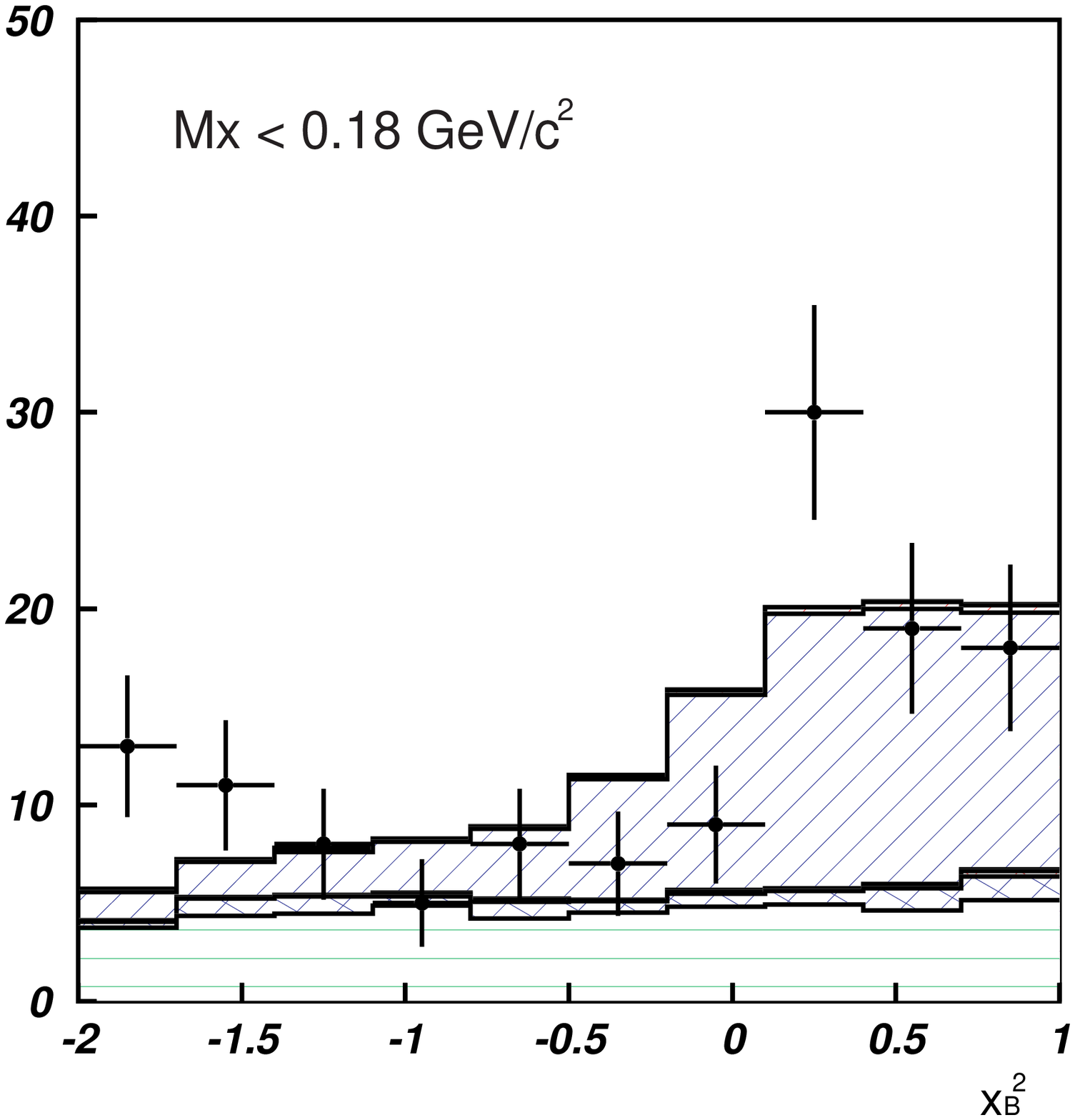,width=2.0in, height=2.0in, angle=0, scale=1.0 }} } &
   \hspace{-0.0cm}{\mbox{\psfig{figure=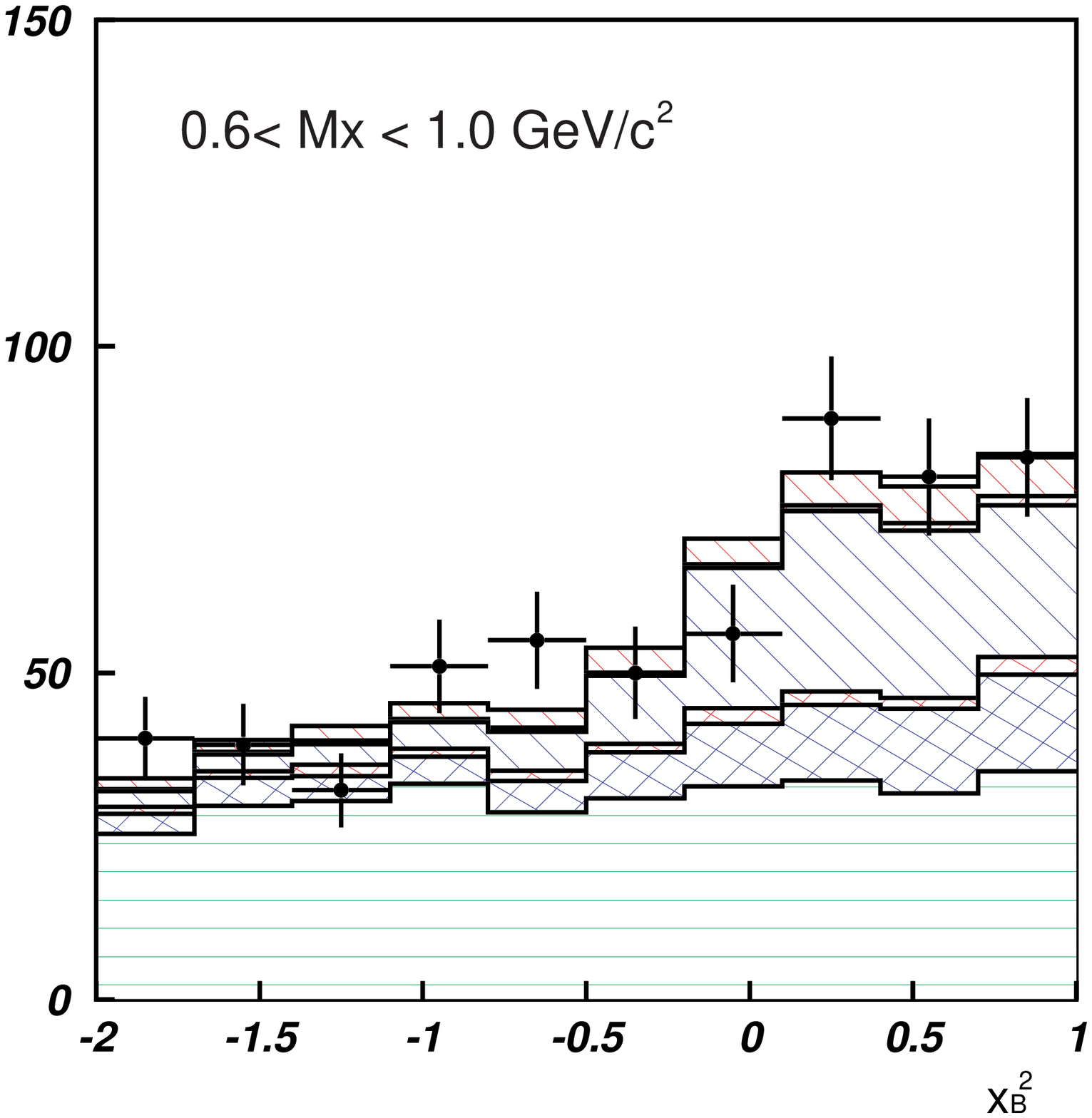,width=2.0in, height=2.0in, angle=0, scale=1.0 }} } \\

  \end{tabular}
  \caption{Projected $M_X$ distribution(left) and $x_B^2$ distributions 
  for the mass region of $\pi$ ($M_X< 0.18 $GeV/$c^2$, middle) 
  and $\rho$ ($ 0.6 < M_X < 1.0 $GeV/$c^2$, right) in all $q^2$ region; points are data.
  Histogram components are $\pi^- \ell^+ \nu$ (red narrow $135^{\circ}$ hatch),
  $\rho^- \ell^+ \nu$ (red wide $45^{\circ}$ hatch), other $X_u \ell^+ \nu$ from $B^0$
  (red cross-hatch) and $\pi^0 \ell^+ \nu$ (blue narrow $45^{\circ}$ hatch),
  $\rho^0 \ell^+ \nu$ (blue wide $135^{\circ}$ hatch), other $X_u \ell^+ \nu$ from $B^+$
  (blue cross-hatch) and $B \bar{B}$ background (green border).}
  \label{fig:fit_allq2}
 \end{center}
\end{figure}

%\subsection{Total Branching Fraction}
%\label{sec:TotalBranch}
Figure~\ref{fig:fit_allq2} presents the projections on $M_X$ and $x_B^2$ of the 
fitting result for data in the entire $q^2$ region.
The extracted yields for the signal components are 
$N(\pi^- \ell^+ \nu) = 150.7 \pm 19.7$, 
$N(\rho^- \ell^+ \nu) = 88.6 \pm 18.9$,
$N(\pi^0 \ell^+ \nu) = 71.7 \pm 11.5$ and 
$N(\rho^0 \ell^+ \nu) = 143.3 \pm 25.4$,
with the LCSR model used for the four signal PDF.

%

\begin{figure}[htbp]
 \begin{center}
  \begin{tabular}{ccc}
   \hspace{-0.0cm}{\mbox{\psfig{figure=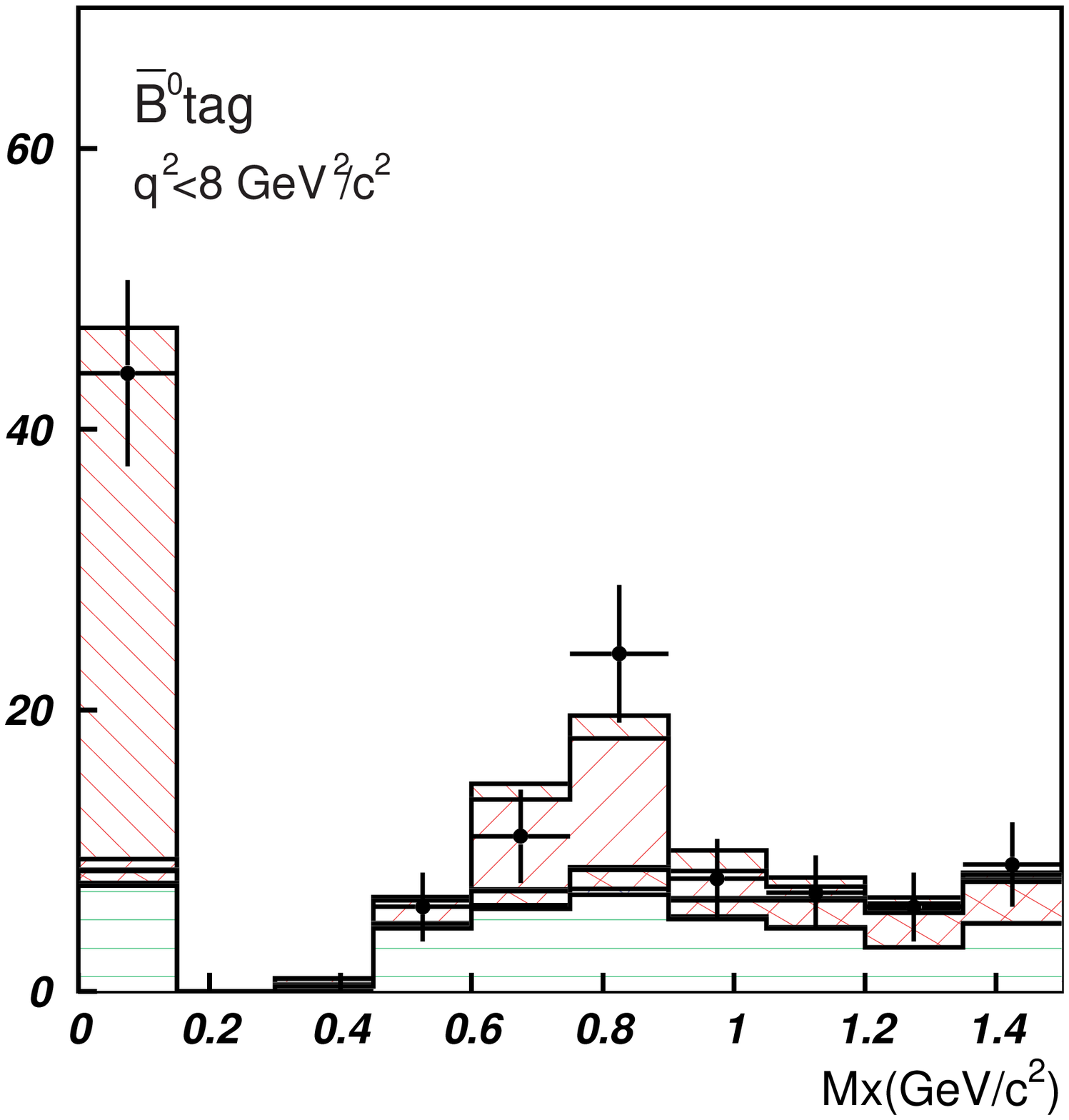,width=2.0in, height=2.0in, angle=0, scale=1.0 }} } &
   \hspace{-0.0cm}{\mbox{\psfig{figure=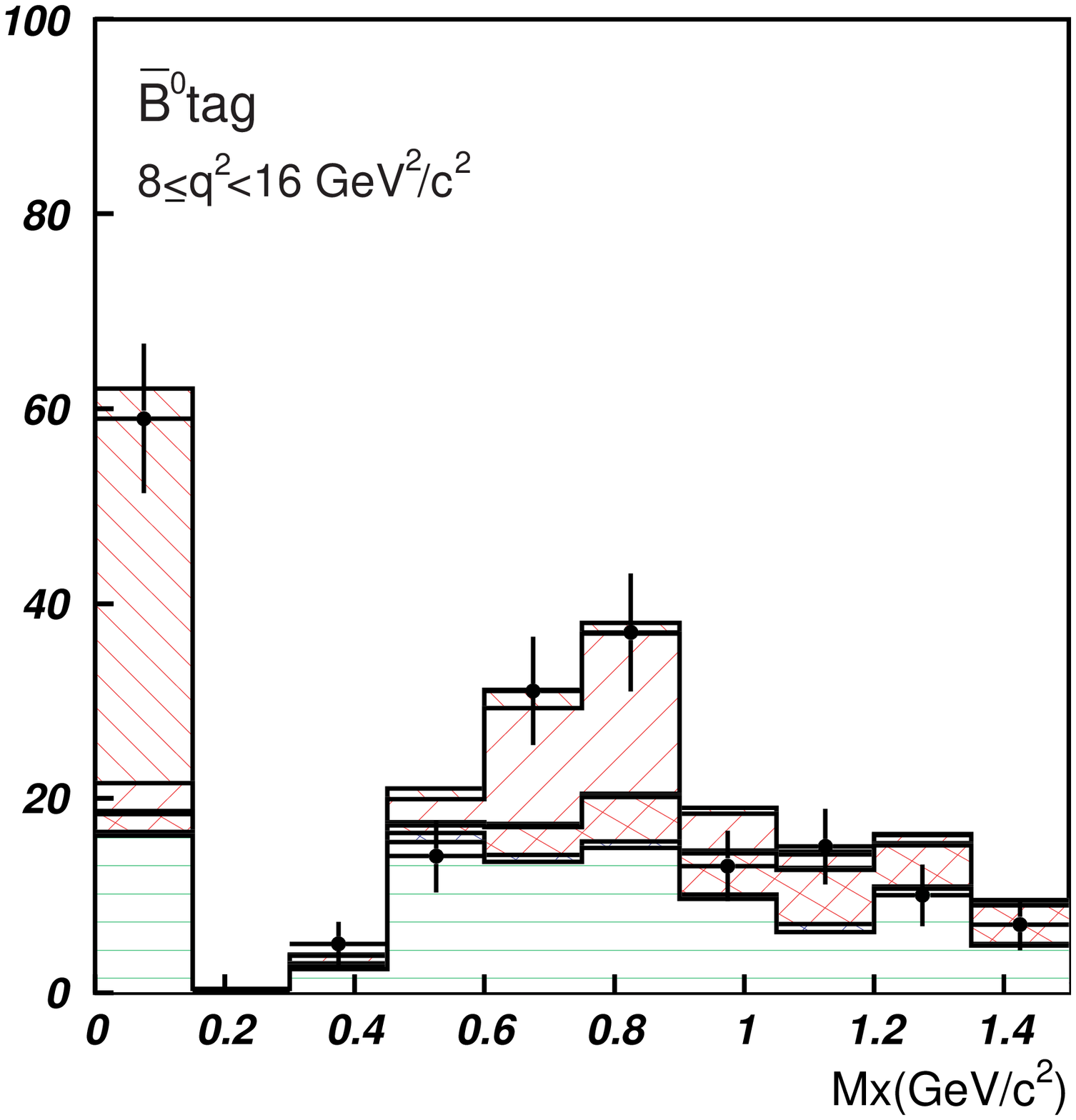,width=2.0in, height=2.0in, angle=0, scale=1.0 }} } &
   \hspace{-0.0cm}{\mbox{\psfig{figure=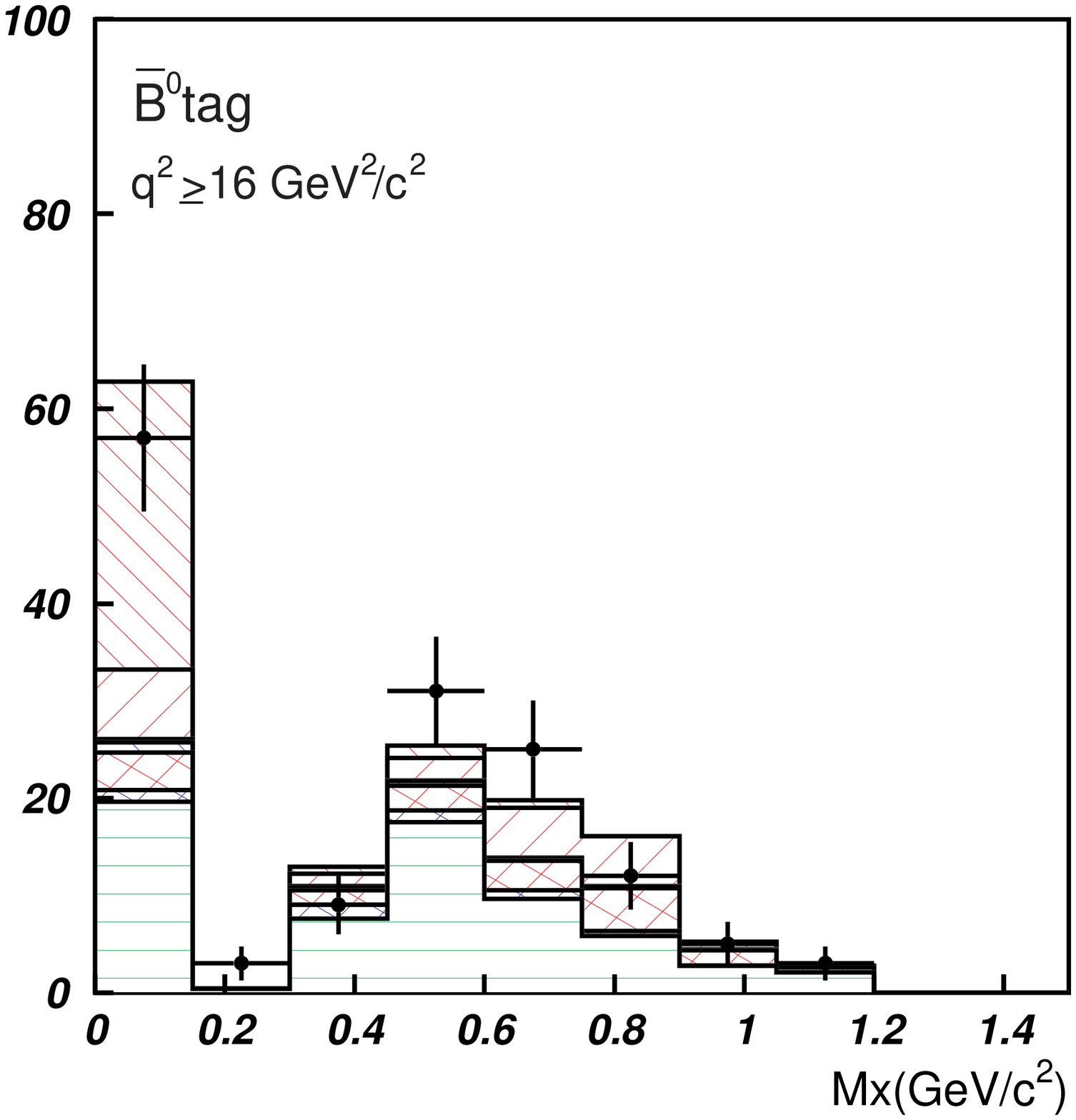,width=2.0in, height=2.0in, angle=0, scale=1.0}} } \\
%   & $B^0 \to \pi^- / \rho^- \ell^+ \nu$& \\
   \hspace{-0.0cm}{\mbox{\psfig{figure=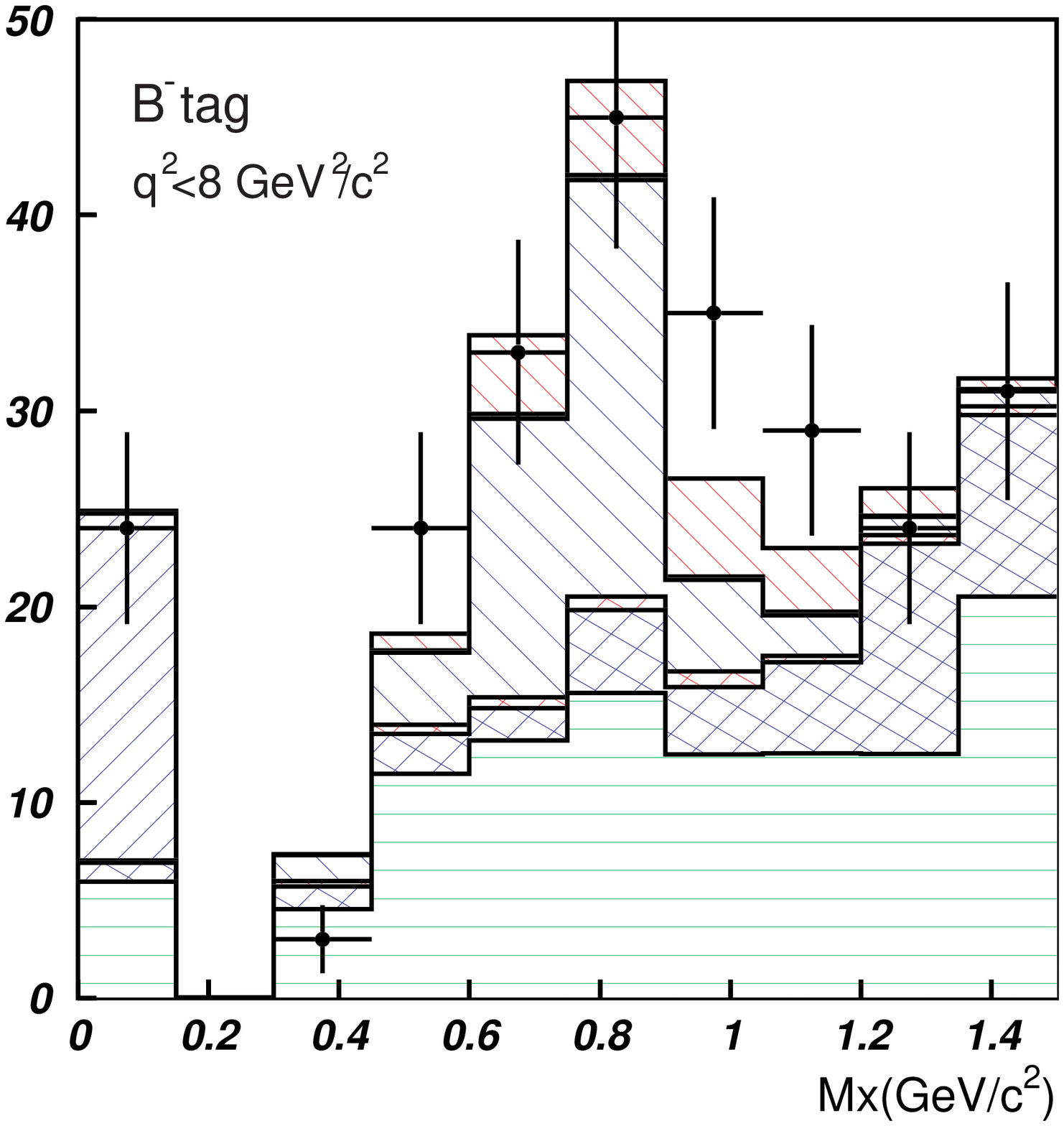,width=2.0in, height=2.0in, angle=0, scale=1.0 }} } &
   \hspace{-0.0cm}{\mbox{\psfig{figure=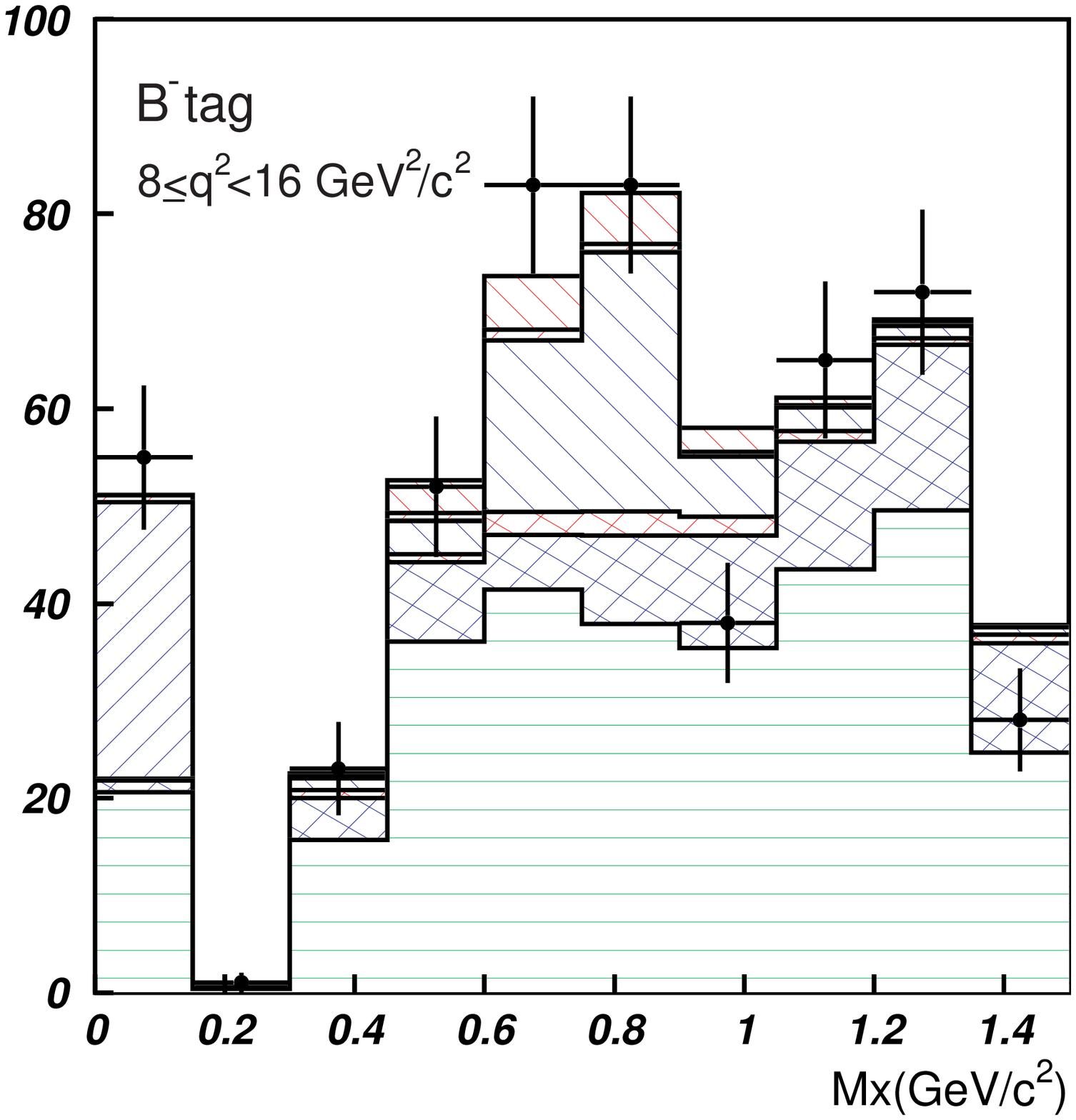,width=2.0in, height=2.0in, angle=0, scale=1.0 }} } &
   \hspace{-0.0cm}{\mbox{\psfig{figure=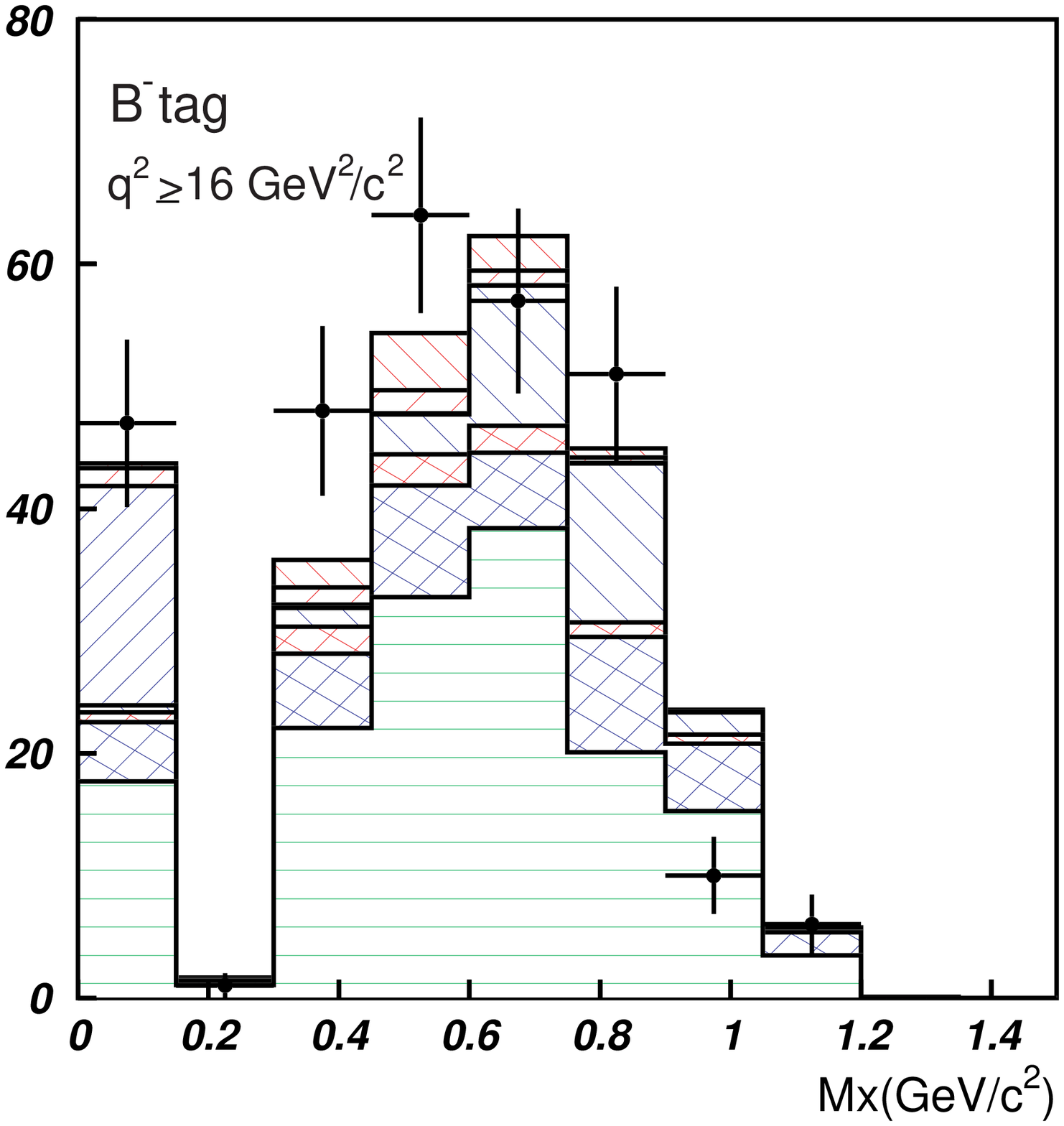,width=2.0in, height=2.0in, angle=0, scale=1.0 } }} \\
%   & $B^+ \to \pi^0 / \rho^0 \ell^+ \nu$& \\
  \end{tabular}
  \caption{Projected $M_X$ distribution in each $q^2$ region;
  Dot is data.   Histogram components are $\pi^- \ell^+ \nu$ (red narrow $135^{\circ}$ hatch),
  $\rho^- \ell^+ \nu$ (red wide $45^{\circ}$ hatch), other $X_u \ell^+ \nu$ from $B^0$
  (red cross-hatch) and $\pi^0 \ell^+ \nu$ (blue narrow $45^{\circ}$ hatch),
  $\rho^0 \ell^+ \nu$ (blue wide $135^{\circ}$ hatch), other $X_u \ell^+ \nu$ from $B^+$
  (blue cross-hatch) and $B \bar{B}$ background (green border).}
  \label{fig:fit_subq2}
 \end{center}
\end{figure}

%\subsection{$q^2$ Distribution}
%\label{sec:q2distribution}
Figure~\ref{fig:fit_subq2} shows projections of the data, separated into three 
$q^2$ bins, $q^2 < 8$ GeV$^2/c^2$ and $8 \leq q^2 < 16$ GeV$^2/c^2$, 
$q^2 \geq 16$ GeV$^2/c^2$.
Here the normalization of the other $B \to X_u \ell \nu$ 
and the $B \bar{B}$ background components are fixed to those obtained in
the above fitting for the entire $q^2$ region.
Table~\ref{tbl:FFdep_q2_Xulnu}
summarizes the extracted branching fractions with different FF-models
for each signal mode, respectively.
The results are unfolded using an efficiency matrix that relates 
the true and reconstructed $q^2$ for the three $q^2$ intervals. 
We calculate the total branching fraction by taking sum of the partial 
branching fractions in the three $q^2$ intervals.

\begin{table}[htbp]
 \begin{center}
  \caption{Extracted branching fractions for each signal mode
    with different FF models in units of 10$^{-4}$; the total branching fraction and the 
    partial branching fractions in three $q^2$ intervals. 
    $\chi^2/n$ and the probability of $\chi^2$ shows the quality of the fit 
    of the FF shape to the extracted $q^2$ distribution.}
  \label{tbl:FFdep_q2_Xulnu}
  \begin{tabular}{l|l|cccccc}
    \hline\hline
Mode & Model    & ${\cal B}_{total}$ & ${\cal B}_{<8}$ & ${\cal B}_{8-16}$ & ${\cal B}_{\geq 16}$ & $\chi^2$ & $Prob.$\\
\hline
$\pi^- \ell^+ \nu$  & Ball'01  & $1.48 \pm 0.20$ & $0.56 \pm 0.11$ & $0.51 \pm 0.11$ & $0.41 \pm 0.12$ & 2.7/3-1 & 0.26 \\
                    & ISGW~II  & $1.45 \pm 0.20$ & $0.53 \pm 0.11$ & $0.53 \pm 0.11$ & $0.39 \pm 0.11$ & 4.2/3-1 & 0.13 \\
                    & UKQCD    & $1.50 \pm 0.20$ & $0.55 \pm 0.11$ & $0.55 \pm 0.12$ & $0.41 \pm 0.11$ & 0.0/3-1 & 0.98 \\
                    & Average  & $1.48 \pm 0.20$ & $0.55 \pm 0.11$ & $0.53 \pm 0.12$ & $0.40 \pm 0.12$ &   --    &  --  \\
\hline
$\rho^- \ell^+ \nu$ & Ball'98  & $2.08 \pm 0.47$ & $0.51 \pm 0.22$ & $1.07 \pm 0.29$ & $0.49 \pm 0.27$ & 2.1/3-1 & 0.35 \\
                    & ISGW~II  & $2.06 \pm 0.47$ & $0.50 \pm 0.20$ & $1.00 \pm 0.28$ & $0.59 \pm 0.30$ & 0.6/3-1 & 0.73 \\
                    & UKQCD    & $1.97 \pm 0.44$ & $0.48 \pm 0.21$ & $1.01 \pm 0.28$ & $0.48 \pm 0.27$ & 1.0/3-1 & 0.59 \\
                    & Melikhov & $2.20 \pm 0.49$ & $0.50 \pm 0.21$ & $1.02 \pm 0.29$ & $0.68 \pm 0.34$ & 3.3/3-1 & 0.19 \\
                    & Average  & $2.07 \pm 0.47$ & $0.50 \pm 0.21$ & $1.02 \pm 0.29$ & $0.56 \pm 0.30$ &   --    &  --  \\
\hline
$\pi^0 \ell^+ \nu$  & Ball'01  & $0.74 \pm 0.13$ & $0.21 \pm 0.06$ & $0.32 \pm 0.08$ & $0.21 \pm 0.08$ & 2.0/3-1 & 0.36 \\
                    & ISGW~II  & $0.76 \pm 0.13$ & $0.20 \pm 0.06$ & $0.33 \pm 0.09$ & $0.22 \pm 0.08$ & 6.6/3-1 & 0.04 \\
                    & UKQCD    & $0.80 \pm 0.14$ & $0.22 \pm 0.07$ & $0.35 \pm 0.09$ & $0.23 \pm 0.08$ & 1.1/3-1 & 0.58 \\
                    & Average  & $0.76 \pm 0.13$ & $0.21 \pm 0.06$ & $0.33 \pm 0.09$ & $0.22 \pm 0.08$ &   --    &  --  \\
\hline
$\rho^0 \ell^+ \nu$ & Ball'98  & $1.40 \pm 0.23$ & $0.50 \pm 0.12$ & $0.57 \pm 0.15$ & $0.33 \pm 0.12$ & 1.5/3-1 & 0.48 \\
                    & ISGW~II  & $1.38 \pm 0.23$ & $0.48 \pm 0.12$ & $0.56 \pm 0.15$ & $0.34 \pm 0.12$ & 3.2/3-1 & 0.21 \\
                    & UKQCD    & $1.36 \pm 0.23$ & $0.48 \pm 0.12$ & $0.58 \pm 0.15$ & $0.31 \pm 0.12$ & 1.4/3-1 & 0.49 \\
                    & Melikhov & $1.40 \pm 0.23$ & $0.51 \pm 0.12$ & $0.57 \pm 0.15$ & $0.32 \pm 0.12$ & 1.9/3-1 & 0.39 \\
                    & Average  & $1.39 \pm 0.23$ & $0.49 \pm 0.12$ & $0.57 \pm 0.15$ & $0.32 \pm 0.12$ &   --    &  --  \\
\hline\hline
  \end{tabular}
 \end{center}
\end{table}

\section{Systematic Errors}
\label{sec:SystematicErrors}
Tables~\ref{tbl:systematic_pi} and {\ref{tbl:systematic_pi0}} summarize the 
experimental systematic errors on the branching fractions.
The experimental systematic errors can be categorized as originating from uncertainties 
in the signal reconstruction efficiency, the background estimation, and
the normalization.
The total experimental systematic error is the quadratic sum of all 
individual ones.
We also consider the systematic error due to the  dependence
of the obtained branching fractions on the FF model.

The effect from the uncertainty on the signal reconstruction efficiency is evaluated based on the efficiency calibration with the 
$B_{sig} \to D^{*-} \ell^+ \nu$ sample, discussed above.
The error is taken to be that on the ratio of observed to expected number of the 
calibration signals (9.3\%).
This gives the largest contribution to the systematic error.
Note this error is dominated by the statistics of the calibration 
signals, as explained above.
Therefore, accumulation of additional integrated luminosity in the future will help to reduce 
this uncertainty.
We further include residual errors for the reconstruction of the signal 
side; 1\% and 2\% for the detection of each charged and neutral pion, 
respectively, and 2\% for the kaon veto and 2.1\% for the lepton selection.

The systematic error due to the uncertainty on the inclusive branching  
fraction ${\cal B}(B \to X_u \ell \nu)$, which is used to constrain 
$X_u^- \ell^+ \nu$ background, is estimated by varying this parameter
within $\pm 1 \sigma$ of the error.
The uncertainty on the $B \bar{B}$ background shape in our selection cut 
($N_{\pi^+}=1$ or $N_{\pi^+}=N_{\pi^0}=1$ for a $\bar{B^0}$ tag and 
 $N_{\pi^0}=1$ or $N_{\pi^+}=N_{\pi^-}=1$ for a $B^-$ tag) is studied in 
the simulation by randomly removing charged tracks and $\pi^0$ according 
to the error in detection efficiency (1\% for a charged track, 2\% for 
$\pi^0$), and also by reassigning identified charged kaons as pions 
according to the uncertainty in the kaon identification efficiency (2\%).
The resultant changes in the extracted branching fractions are assigned
as systematic errors. 
We have seen significant change for $\rho \ell^+ \nu$ in the high $q^2$
region ($q^2 > 16$~GeV$^2/c^2$).
We also vary the fraction of $B \to D^{**} \ell \nu$ decays in the
$B\bar{B}$ background MC by the error quoted in~\cite{PDG2005} to test 
the $B \to X_c \ell \nu$ model dependence in the $B\bar{B}$ background shape.
To assess the uncertainty due to the production rate of $K_L^0$, we vary
the production rate by the error of the inclusive branching fraction of 
$B \to K^0 ~X$ quoted in~\cite{PDG2005}.
  
As for the normalization, we consider the uncertainty in the number of 
$B^0 \bar{B^0}$ and $B^+ B^-$ pairs; 
the ratio of $B^+B^-$ to $B^0 \bar{B^0}$ pairs ($f_+/f_0$,
2.5\%), the mixing parameter ($\chi_d$, 1.0\%) and the measured
number of $B \bar{B}$ pairs ($N_{B\bar{B}}$, 1.1\%).

\begin{table}[htbp]
 \begin{center}
  \caption{Summary of systematic errors(\%) for $B^0 \to \pi^- / 
\rho^- \ell^+ \nu$.}
  \label{tbl:systematic_pi}

  \begin{tabular}{c|ccc|c|c|ccc|c|c}
   \hline\hline
          & \multicolumn{5}{c|}{$\pi^{-} \ell^{+} \nu$} & \multicolumn{5}{c}{$\rho^{-} \ell^{+} \nu$}            \\
   Source & $q^2 < 8$ & $ 8 - 16$ & $\geq 16$ & $< 16$ & all & $q^2 < 8$ & $ 8 - 16$ & $\geq 16$ & $< 16$ & all     \\
   \hline
   Tracking efficiency        &      1  &    1  &    1  &    1  &    1  &     1  &    1  &    1  &    1  &    1  \\ 
   $\pi^{0}$ reconstruction   &     --  &   --  &   --  &   --  &   --  &     2  &    2  &    2  &    2  &    2  \\ 
   Lepton identification      &    2.1  &  2.1  &  2.1  &  2.1  &  2.1  &   2.1  &  2.1  &  2.1  &  2.1  &  2.1  \\ 
   Kaon identification        &      2  &    2  &    2  &    2  &    2  &     2  &    2  &    2  &    2  &    2  \\ 
   $D^* \ell \nu$ calibration &    9.3  &  9.3  &  9.3  &  9.3  &  9.3  &   9.3  &  9.3  &  9.3  &  9.3  &  9.3  \\ 
   $Br(X_u \ell \nu)$ in the fitting        
                              &    0.2  &  2.0  &  2.1  &  1.1  &  1.4  &   0.8  &  4.2  & 13.1  &  1.1  &  5.4  \\ 
   $B\bar{B}$ background shape
                              &    1.5  &  1.7  &  3.5  &  1.3  &  1.7  &   2.5  &  1.8  &  9.0  &  1.3  &  2.6  \\
   $Br(D^{**} \ell \nu)$      &    1.2  &  1.4  &  0.1  &  1.4  &  0.9  &   0.5  &  0.1  &  1.3  &  1.4  &  0.4  \\
   $K_L^0$ production rate    &    0.2  &  0.3  &  0.4  &  0.3  &  0.3  &   1.0  &  0.8  &  2.2  &  0.3  &  1.2  \\
   $N_{B\overline{B}}$        &    1.1  &  1.1  &  1.1  &  1.1  &  1.1  &   1.1  &  1.1  &  1.1  &  1.1  &  1.1  \\ 
   $\it{f}_+ / \it{f}_0$      &    2.5  &  2.5  &  2.5  &  2.5  &  2.5  &   2.5  &  2.5  &  2.5  &  2.5  &  2.5  \\ 
   $\chi_d$                   &    1.1  &  1.1  &  1.1  &  1.1  &  1.1  &   1.1  &  1.1  &  1.1  &  1.1  &  1.1  \\ 
   \hline
   exp. total                 &   10.4  & 10.7  & 11.0  &  10.5 &  10.5 &  10.8  & 11.4  & 19.2  & 10.5  & 12.1  \\ 
   \hline\hline
   F.F for signal             &    2.6  &  4.0  &  3.1  &  1.9  &  1.7  &   3.3  &  5.4  & 21.4  &  1.9  &  6.2  \\ 
   F.F for cross-feed         &    2.2  &  3.9  &  5.1  &  2.8  &  1.9  &   2.8  &  2.0  &  6.6  &  2.8  &  2.5  \\ 
   \hline
   F.F total                  &    3.4  &  5.6  &  6.0  &  3.3  &  2.5  &   4.3  &  5.8  & 22.4  &  4.8  &  6.7  \\
   \hline\hline
  \end{tabular}
 \end{center}
\end{table}

\begin{table}[htbp]
 \begin{center}
  \caption{Summary of systematic errors(\%) for $B^+ \to \pi^0 / 
\rho^0 \ell^+ \nu$.}
  \label{tbl:systematic_pi0}

  \begin{tabular}{c|ccc|c|c|ccc|c|c}
   \hline\hline
          & \multicolumn{5}{c|}{$\pi^0 \ell^{+} \nu$} & \multicolumn{5}{c}{$\rho^0 \ell^{+} \nu$}            \\
   Source & $q^2 < 8$ & $ 8 - 16$ & $\geq 16$ & $< 16$ & all & $q^2 < 8$ & $ 8 - 16$ & $\geq 16$ & $< 16$ & all     \\
   \hline
   Tracking efficiency        &     --  &   --  &   --  &   --  &   --  &     2  &    2  &    2  &    2  &    2  \\ 
   $\pi^{0}$ reconstruction   &      2  &    2  &    2  &    2  &    2  &    --  &   --  &   --  &   --  &   --  \\ 
   Lepton identification      &    2.1  &  2.1  &  2.1  &  2.1  &  2.1  &   2.1  &  2.1  &  2.1  &  2.1  &  2.1  \\ 
   Kaon identification        &     --  &   --  &   --  &   --  &   --  &     4  &    4  &    4  &    4  &    4  \\ 
   $D^* \ell \nu$ calibration &    9.3  &  9.3  &  9.3  &  9.3  &  9.3  &   9.3  &  9.3  &  9.3  &  9.3  &  9.3  \\ 
   $Br(X_u \ell \nu)$ in the fitting        
                              &    0.4  &  3.7  &  0.5  &  1.3  &  1.6  &   2.3  &  2.5  & 10.8  &  2.5  &  4.4  \\ 
   $B\bar{B}$ background shape
                              &    2.2  &  5.6  &  2.7  &  2.3  &  3.8  &   5.9  &  4.3  & 12.9  &  1.7  &  1.9  \\
   $Br(D^{**} \ell \nu)$      &    1.3  &  0.7  &  0.8  &  0.5  &  0.9  &   0.2  &  1.7  &  3.1  &  0.9  &  1.4  \\
   $K_L^0$ production rate    &    0.4  &  1.1  &  0.7  &  0.5  &  0.8  &   0.5  &  1.1  &  1.8  &  0.2  &  0.3  \\
   $N_{B\overline{B}}$        &    1.1  &  1.1  &  1.1  &  1.1  &  1.1  &   1.1  &  1.1  &  1.1  &  1.1  &  1.1  \\ 
   $\it{f}_+ / \it{f}_0$      &    2.5  &  2.5  &  2.5  &  2.5  &  2.5  &   2.5  &  2.5  &  2.5  &  2.5  &  2.5  \\ 
   \hline
   exp. total                 &   10.5  & 12.2  & 10.5  & 10.5  & 11.0  &  12.6  & 12.1  & 20.5  & 11.3  & 12.0  \\ 
   \hline\hline
   F.F for signal             &    5.1  &  4.5  &  4.4  &  4.7  &  4.4  &   2.8  &  1.3  &  6.1  &  1.7  &  1.8  \\ 
   F.F for cross-feed         &    0.1  &  2.1  &  2.6  &  1.6  &  1.4  &   1.0  &  1.3  &  4.4  &  1.1  &  0.1  \\ 
   \hline
   F.F total                  &    5.1  &  5.0  &  5.1  &  5.0  &  4.6  &   2.9  &  1.8  &  7.5  &  2.0  &  1.8  \\
   \hline\hline
  \end{tabular}
 \end{center}
\end{table}

%\subsection{FF Model Dependence and Summary of the Measurement}
%\label{sec:FFdependence}
The dependence of the extracted branching fractions on the FF model has been 
studied by repeating the above fitting procedure with various FF 
models for the signal mode and also for the cross-feed mode 
($\pi \ell \nu \leftrightarrow \rho \ell \nu$).
We consider the models listed in Table~\ref{tbl:FFdep_q2_Xulnu}.
For the extracted ${\cal B}(B \to \pi^-(\pi^0) \ell^+ \nu)$,
the standard deviation among the models is $<1.7(4.4)$\% for 
$\pi \ell^+ \nu$  and $<1.9(1.4)$\% for  
$\rho \ell^+ \nu$.
For ${\cal B}(B \to \rho^-(\rho^0) \ell^+ \nu)$, the standard deviation is
$<6.2(1.8)$\% for 
$\rho \ell^+ \nu$ and $<2.5(0.1)$\% for  
$\pi \ell^+ \nu$.
The assigned total error due to the FF model dependence is the
quadratic sum of the maximum variations with the signal and cross-feed 
FF models.

\section{Results}
Table~\ref{tbl:summary_br} summarizes our measurements of the total and partial 
branching fractions for the four signal modes.
Each branching fraction is obtained by taking the simple average of 
the values obtained from the FF models shown in 
Table~\ref{tbl:FFdep_q2_Xulnu}.
The errors shown in the table are statistical, experimental systematic,
and systematic due to form-factor uncertainties.
The obtained branching fractions for $B^0 \to \pi^-/\rho^- \ell^+ \nu$
are consistent with the existing measurements by CLEO~\cite{CLEO2003} 
and BaBar~\cite{BABAR2003,BABAR2005,BABAR_SLTAG1,BABAR_FLTAG}, within 
the measurement uncertainties.
The overall uncertainty on our result for $B^0 \to \pi^- \ell^+ \nu$ is 18\%,  comparable to that on the measurement published by CLEO~\cite{CLEO2003} which is based on $\nu$-reconstruction.

\begin{table}[htbp]
 \begin{center}
  \caption{Summary of the obtained branching fractions.
The errors are statistical, experimental systematic, and systematic due 
to form-factor uncertainties.}
  \label{tbl:summary_br}
  \begin{tabular}{cc|c}
   \hline\hline
   Modes    &   $q^2$ region (GeV$^2/c^2)$  & Branching fraction ($\times 10^{-4}$)\\
   \hline
   $B^0 \to \pi^- \ell^+ \nu$   & ~Total    & $1.48 \pm 0.20 \pm 0.16 \pm 0.04$ \\
                                & ~$ > 16$  & $0.40 \pm 0.12 \pm 0.04 \pm 0.02$ \\
                                & ~$ < 16$  & $1.08 \pm 0.16 \pm 0.11 \pm 0.04$ \\
   \hline
   $B^+ \to \pi^0 \ell^+ \nu$   & ~Total    & $0.76 \pm 0.13 \pm 0.08 \pm 0.04$ \\
                                & ~$ > 16$  & $0.22 \pm 0.08 \pm 0.02 \pm 0.01$ \\
                                & ~$ < 16$  & $0.54 \pm 0.11 \pm 0.06 \pm 0.03$ \\
   \hline
   $B^0 \to \rho^- \ell^+ \nu$  & ~Total    & $2.07 \pm 0.47 \pm 0.25 \pm 0.14$ \\   
   \hline
   $B^+ \to \rho^0 \ell^+ \nu$  & ~Total    & $1.39 \pm 0.23 \pm 0.17 \pm 0.02$ \\   
   \hline\hline
  \end{tabular}
 \end{center}
\end{table}

\begin{figure}[htbp]
 \begin{center}
  \begin{tabular}{cc}
   \hspace{-0.0cm}
   \mbox{\psfig{figure=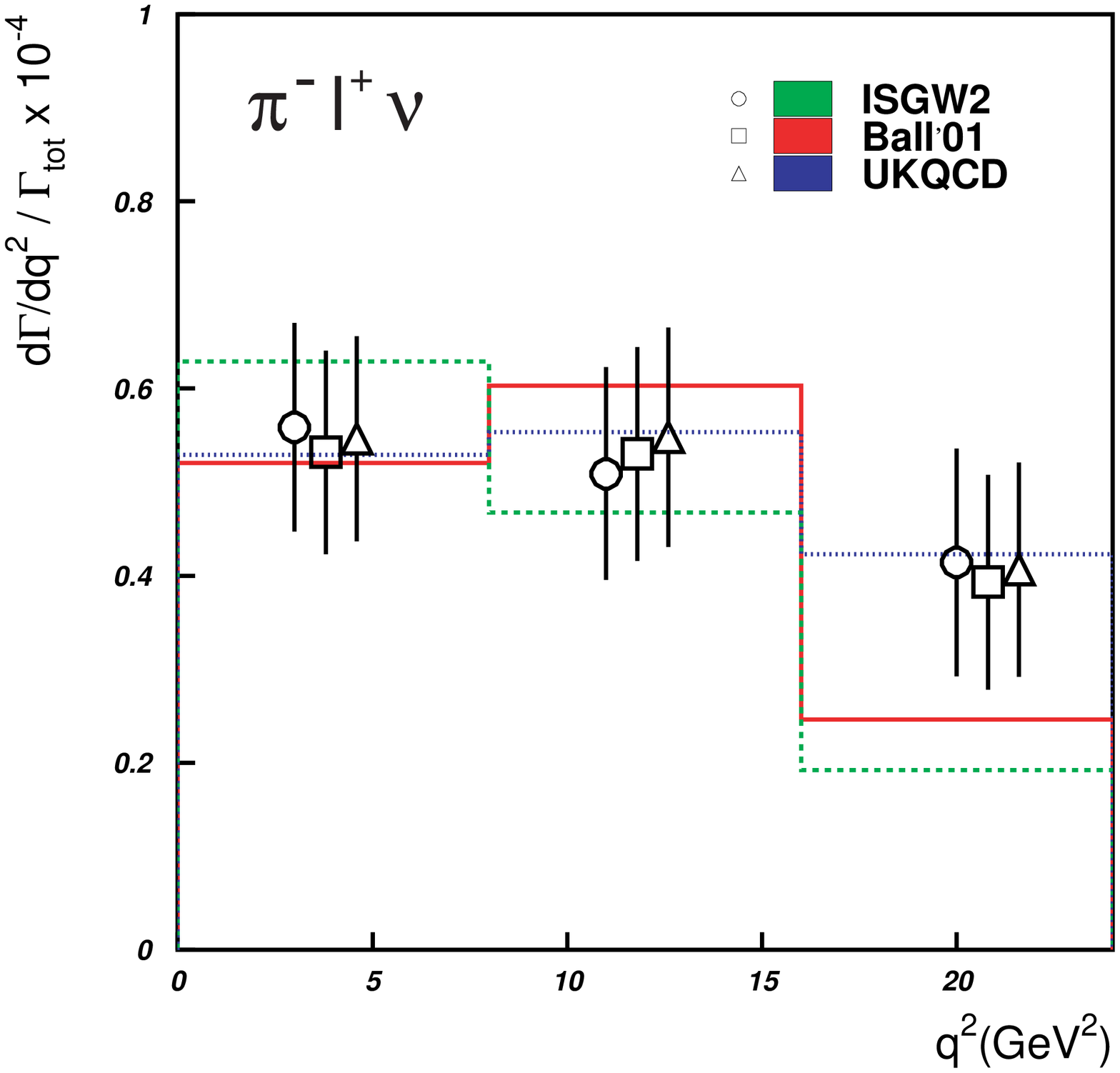,width=2.5in, height=2.5in, angle=0, scale=1.0 } } &
   \hspace{-0.0cm}{\mbox{\psfig{figure=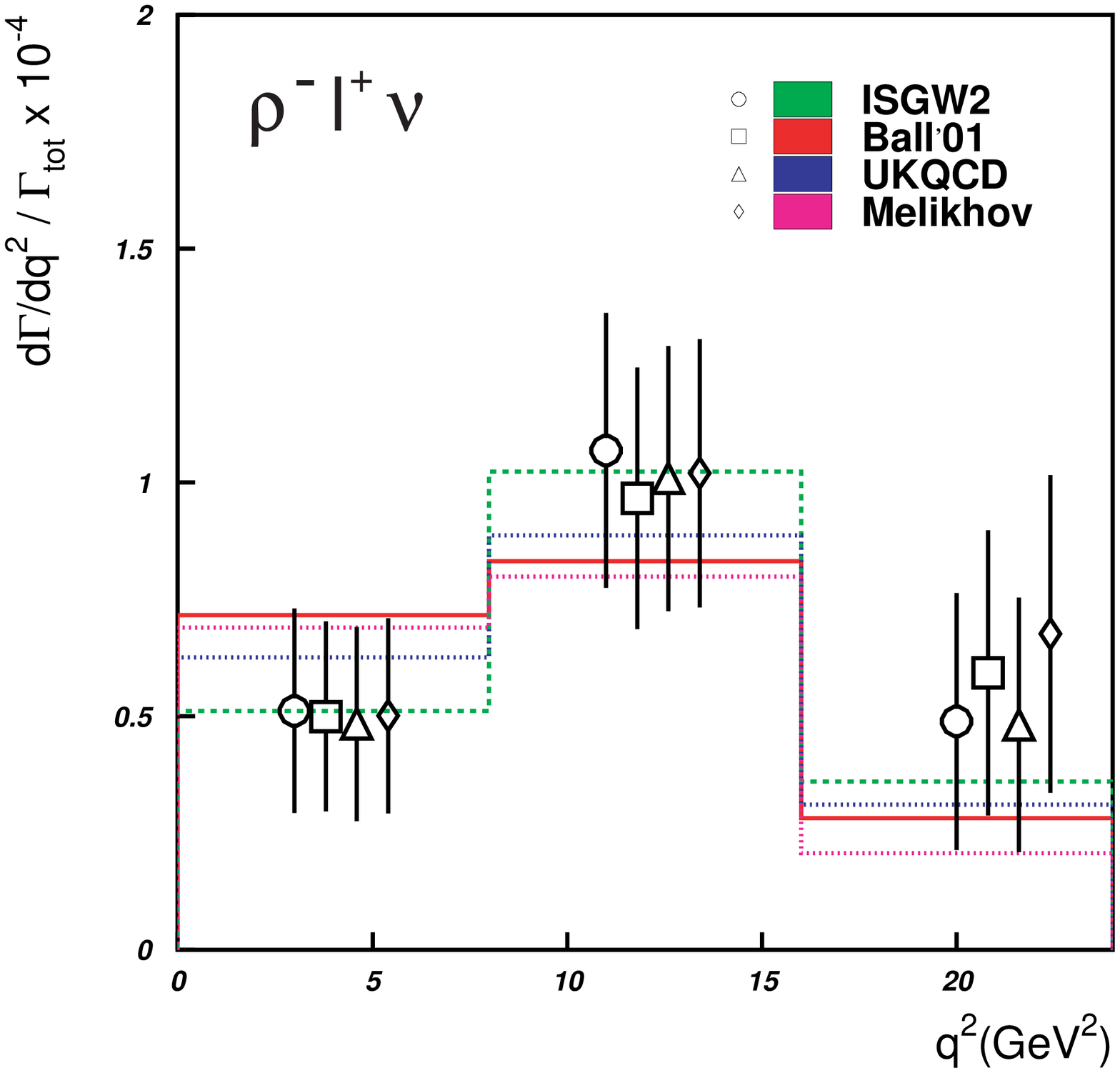,width=2.5in, height=2.5in, angle=0, scale=1.0 } }} \\
%   $\pi^- \ell^+ \nu$   &   $\rho^- \ell^+ \nu$   \\
   \hspace{-0.0cm}{\mbox{\psfig{figure=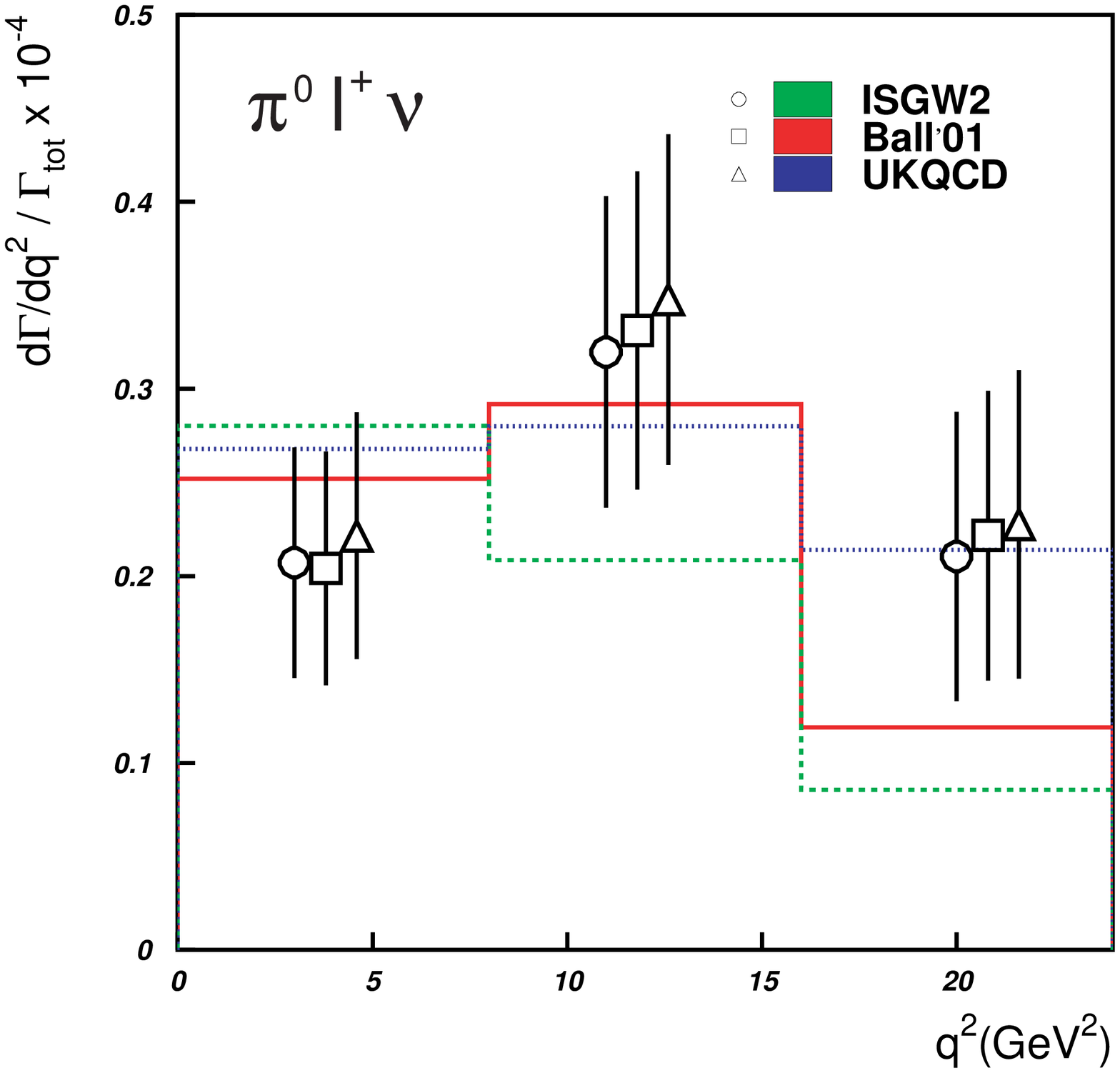,width=2.5in, height=2.5in, angle=0, scale=1.0 } }} &
   \hspace{-0.0cm}{\mbox{\psfig{figure=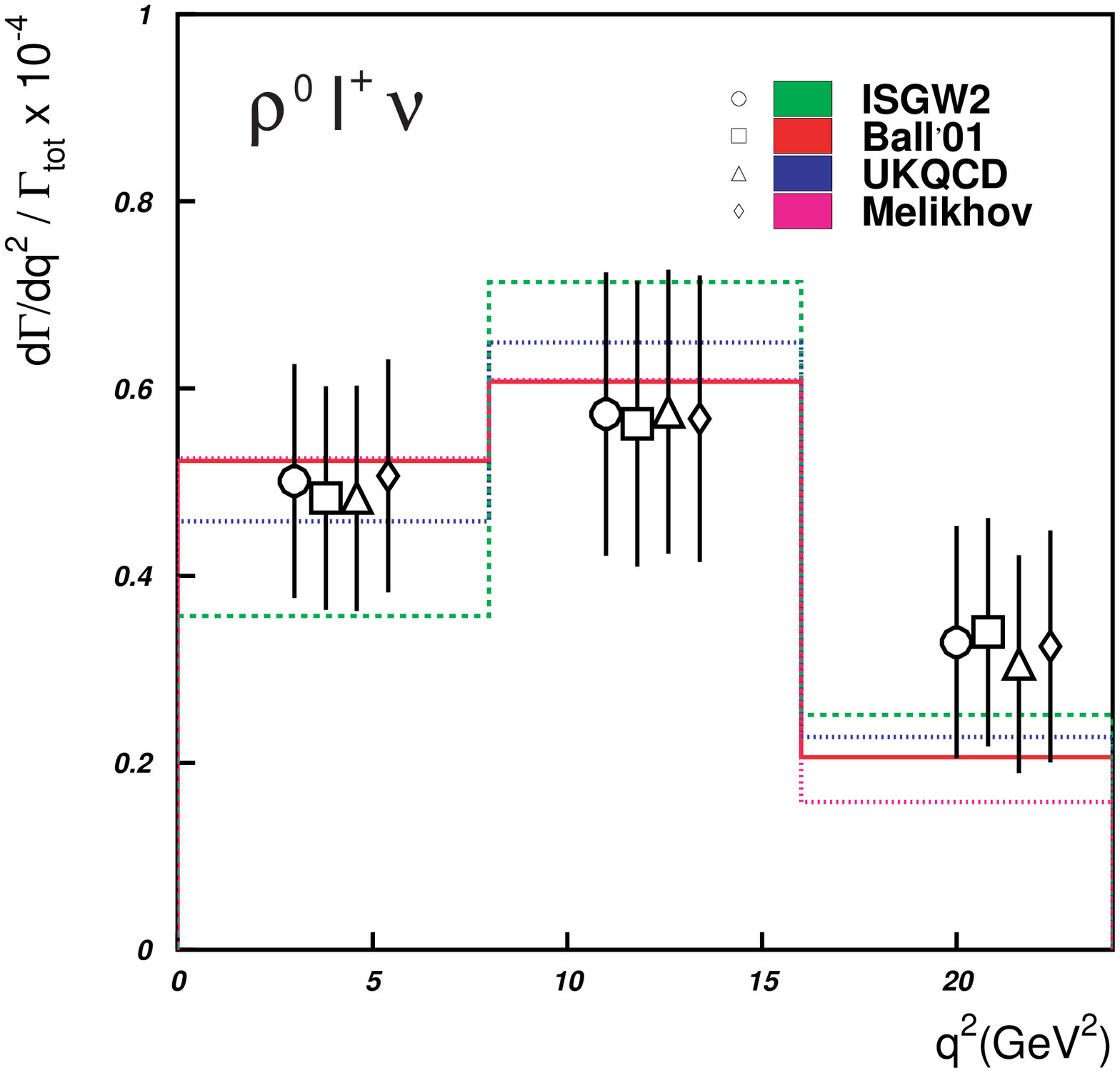,width=2.5in, height=2.5in, angle=0, scale=1.0 } }} \\
%   $\pi^0 \ell^+ \nu$   &      $\rho^0 \ell^+ \nu$

  \end{tabular}
  \caption{Extracted $q^2$ distribution. Data points are shown for different FF
    models used to estimate the detection efficiency. 
    Lines are for the best fit of the FF shapes to the obtained $q^2$ 
    distribution.}
  \label{fig:q2dist}
 \end{center}
\end{figure}

Figure~\ref{fig:q2dist} presents the measured $q^2$ distributions for 
each signal mode, overlaid with the best fits of FF shapes to the data.
To be self-consistent, the shape of a particular FF model is fit to the 
$q^2$ distribution extracted with the same FF model.
The quality of the fit in terms of $\chi^2$ and the probability of 
$\chi^2$, shown in Table~\ref{tbl:FFdep_q2_Xulnu}, may provide one way 
to discriminate among the models.
For both $B^0 \to \pi^- \ell^+ \nu$ and $B^+ \to \pi^0 \ell^+ \nu$, the 
agreement with the ISGW II model is marginal, as indicated by other 
measurements~\cite{CLEO2003, BABAR2005}.

In this work, the $B^0 \to \pi^- \ell^+ \nu / B^+ \to \pi^0 \ell^+ \nu$
and $B^0 \to \rho^- \ell^+ \nu / B^+ \to \rho^0 \ell^+ \nu$ signals
are extracted separately, which allows us to test the isospin relations.
>From the obtained branching fractions and the $B$ meson lifetime in
~\cite{PDG2005}, the ratios of decay rates are found to be,
\begin{eqnarray}
\frac{\Gamma(B^0 \to \pi^- \ell^+ \nu)}{\Gamma(B^+ \to \pi^0 \ell^+ \nu)}
= (2.08 \pm 0.45 \pm 0.17),\\
\frac{\Gamma(B^0 \to \rho^- \ell^+ \nu)}{\Gamma(B^+ \to \rho^0 \ell^+ \nu)}
= (1.59 \pm 0.44 \pm 0.17),
\end{eqnarray}  
where the first and second errors are statistical and systematic errors,
respectively.
The both ratios are found to be consistent with the isospin relations;
$\Gamma(B^0 \to \pi^-(\rho^-) \ell^+ \nu) = 
2\Gamma(B^+ \to \pi^0(\rho^0) \ell^+ \nu)$. 

%
%\section{$|V_{ub}|$ Determination}
%
The obtained branching fractions in Table~\ref{tbl:summary_br}
can be used to extract $|V_{ub}|$ using the relation,
\begin{equation}
|V_{ub}| = \sqrt{ \frac{{\cal B}(B \to \pi \ell^+ \nu) }
  {{\tilde \Gamma}_{thy}~\tau_{B} } },
\end{equation}
where ${\tilde \Gamma}_{thy}$ is the form-factor normalization, 
predicted from theories.
We list the partial branching fractions for $B \to \pi \ell^+ \nu$
decays in the $q^2$ region above 16 GeV$^2/c^2$, where the LQCD 
calculations are the most sensitive.
The table provides also the results in the region below 
16 GeV$^2/c^2$, so that one can deduce $|V_{ub}|$ based on other
approaches such as the LCSR calculations~\cite{Ball2005}.

In this paper, we quote $|V_{ub}|$ based on the $\pi \ell^+ \nu$ data 
in the high $q^2$ region and the form factor predicted by recent 
unquenched LQCD calculations. 
Their predicted ${\tilde \Gamma}_{thy}$ for the $q^2 \geq 16$\,GeV$^2/c^2$ 
region are ${\tilde \Gamma}_{thy} = 1.31 \pm 0.33$ ps$^{-1}$ (HPQCD)
~\cite{Shigemitsu} and $1.83 \pm 0.50$ ps$^{-1}$ (FNAL)~\cite{Okamoto}.
We use $\tau_{B^0} = 1.532\pm 0.009$ ps and $\tau_{B^+} = 1.638\pm 0.011$ ps 
~\cite{PDG2005}, and we assume the isospin relations 
for ${\tilde \Gamma}_{thy}$ between $B^0 \to \pi^-$ and $B^+ \to \pi^0$
transitions. 
Table~\ref{tbl:summary_vub} summarize the results, where the first and
second errors are the experimental statistical and systematic errors,
respectively.
The third error is based on the error on
${\tilde \Gamma}_{thy}$ quoted by the LQCD authors.
These theoretical errors are asymmetric because we assign them by taking the 
variation in $|V_{ub}|$ when ${\tilde \Gamma}_{thy}$ is varied by the 
quoted errors.  
By taking the average of the results obtained from $B^0 \to \pi^- \ell^+ \nu$ 
and $B^+ \to \pi^0 \ell^+ \nu$ data, we obtain
\begin{eqnarray}
|V_{ub}|_{(q^2 \geq 16)}^{\pi^- \ell^+ \nu + \pi^0 \ell^+ \nu } & = &
(4.50 \pm 0.52 \pm 0.27 ^{+0.70}_{-0.48}) \times 10^{-3} (\mbox{HPQCD}), \\
|V_{ub}|_{(q^2 \geq 16)}^{\pi^- \ell^+ \nu + \pi^0 \ell^+ \nu } & = &
(3.81 \pm 0.44 \pm 0.23 ^{+0.66}_{-0.43}) \times 10^{-3} (\mbox{FNAL}).
\end{eqnarray}

\begin{table}[htbp]
 \begin{center}
  \caption{Summary of $|V_{ub}|$ obtained from the $B \to \pi \ell^+ \nu$
data in the $q^2 > 16$\,GeV$^2/c^2$ region. 
The first and second errors are experimental statistical and systematic errors, 
respectively. The third error stems from the error on
${\tilde \Gamma}_{thy}$ quoted by the LQCD authors.}
  \label{tbl:summary_vub}
  \begin{tabular}{cccc}
   \hline\hline
  Theory  &  ${\tilde \Gamma}_{thy}$(ps$^{-1}$) & Mode & $|V_{ub}| (\times 10^{-3})$ \\
   \hline
  HPQCD   &  $1.31 \pm 0.33$ & $\pi^- \ell^+ \nu$                       & $4.49 \pm 0.65 \pm 0.28 ^{+0.70}_{-0.48}$ \\
          &                  & $\pi^0 \ell^+ \nu$                       & $4.53 \pm 0.81 \pm 0.26 ^{+0.71}_{-0.48}$ \\
          &                  & $\pi^- \ell^+ \nu + \pi^0 \ell^+ \nu$    & $4.50 \pm 0.52 \pm 0.27 ^{+0.70}_{-0.48}$ \\
   \hline
  FNAL    &  $1.83 \pm 0.50$ & $\pi^- \ell^+ \nu$                       & $3.80 \pm 0.55 \pm 0.24 ^{+0.66}_{-0.43}$ \\
          &                  & $\pi^0 \ell^+ \nu$                       & $3.83 \pm 0.69 \pm 0.22 ^{+0.66}_{-0.44}$ \\
          &                  & $\pi^- \ell^+ \nu + \pi^0 \ell^+ \nu$    & $3.81 \pm 0.44 \pm 0.23 ^{+0.66}_{-0.43}$ \\
   \hline\hline
  \end{tabular}
 \end{center}
\end{table}

These values
are in agreement with those from the inclusive $B \to X_u \ell \nu$ decays
~\cite{HFAG_LP05}.
The experimental precision in the above $|V_{ub}|$ determination 
is 13\%, and it is currently dominated by the statistical error of 12\%.
By accumulating more luminosity, a measurement with errors below 10\% is
feasible.
With improvement in unquenched LQCD calculations, the present measurement 
will provide a more accurate determination of $|V_{ub}|$.

%\begin{figure}[htbp]
%%\vspace{-0.7cm}
% \begin{center}
%  \mbox{\psfig{figure=./fig/Fig6.eps,width=5.0in, height=5.0in, angle=0, scale=0.9 } } 
%  \caption{ $|V_{ub}|$ determined using LQCD and the $\pi^- \ell^+ \nu$ 
%    branching fraction in the $q^2 \geq 16$\,GeV$^2/c^2$ region.
%    For each data point, $\pm 1 \sigma$ range of the experimental 
%    errors (quadratic sum of statistical and systematic errors) 
%    and the range of the total error including the theoretical error are indicated. 
%    The CLEO result~\cite{CLEO2003} with the averaged quenched LQCD formfactor is also shown for comparison.}
%  \label{fig:pilnu_lqcd}
% \end{center}
%\end{figure}

%\begin{figure}[htbp]
%%\vspace{-0.7cm}
% \begin{center}
%  \mbox{\psfig{figure=./fig/Fig7.eps,width=5.0in, height=5.0in, angle=0, scale=0.9 } } 
%  \caption{ Summary of Belle $|V_{ub}|$ determination.
%            Results based on the exclusive $B^0 \to \pi^- \ell^+ \nu$ measurement
%            ($q^2 \geq 16$GeV$^2/c^2$) in this work with the averaged queched LQCD
%            form factor (top) and with preliminary unqueched LQCD form factors
%            (middle two) are shown, together with those based on our inclusive
%            $B \to X_u \ell \nu$ measurements (bottom three). The lower two inclusive
%             results are rescaled using the recent determination of the $f_u$ factor~\cite{limosani}.}
%  \label{fig:belle_vub}
% \end{center}
%\end{figure}

\vspace{1.0cm}
%\section*{Acknowledgments}
%***** Acknowledgments *****
% Please paste this acknowledgement into your latex file. 
 %----------- Long version, for most papers ----------- 

We thank the KEKB group for the excellent operation of the
accelerator, the KEK Cryogenics group for the efficient
operation of the solenoid, and the KEK computer group and
the National Institute of Informatics for valuable computing
and Super-SINET network support. We acknowledge support from
the Ministry of Education, Culture, Sports, Science, and
Technology of Japan and the Japan Society for the Promotion
of Science; the Australian Research Council and the
Australian Department of Education, Science and Training;
the National Science Foundation of China under contract
No.~10175071; the Department of Science and Technology of
India; the BK21 program of the Ministry of Education of
Korea and the CHEP SRC program of the Korea Science and
Engineering Foundation; the Polish State Committee for
Scientific Research under contract No.~2P03B 01324; the
Ministry of Science and Technology of the Russian
Federation; the Ministry of Education, Science and Sport of
the Republic of Slovenia; the National Science Council and
the Ministry of Education of Taiwan; and the U.S.\
Department of Energy.

%-------- Short version, if necessary, for PRL -----------
% currently commented out
%We thank the KEKB group for the excellent operation of the
%accelerator, the KEK Cryogenics group for the efficient
%operation of the solenoid, and the KEK computer group and
%the NII for valuable computing and Super-SINET network
%support.  We acknowledge support from MEXT and JSPS (Japan);
%ARC and DEST (Australia); NSFC (contract No.~10175071,
%China); DST (India); the BK21 program of MOEHRD and the CHEP
%SRC program of KOSEF (Korea); KBN (contract No.~2P03B 01324,
%Poland); MIST (Russia); MESS (Slovenia); NSC and MOE
%(Taiwan); and DOE (USA).

%

\end{document}